\documentstyle[aps,amssymb,pre,multicol,epsfig]{revtex}
\begin{document}

\draft    
\author{I.V. Barashenkov \cite{add:igor},
E.V. Zemlyanaya
\cite{add:elena},
M. B\"ar
\cite{email:markus}}
\address{Max-Planck-Institut f\"ur Physik
komplexer
Systeme, N\"othnitzer Str.38, Dresden, Germany}

\title{Travelling solitons in the parametrically
driven nonlinear Schr\"odinger equation}

\maketitle
\begin{abstract}
We show that
 the  parametrically driven nonlinear
Schr\"odinger equation
 has wide classes of travelling soliton solutions,
some of which are stable.
For small driving strengths stable nonpropagating and moving solitons
co-exist while  strongly forced solitons can only be stable
when moving sufficiently fast.

\end{abstract}


\begin{multicols}{2}
\section{Introduction}
The parametrically driven damped nonlinear Schr\"odinger
(NLS) equation,
\begin{equation}
\label{NLS}
i \psi_t +\psi_{XX} +2 |\psi|^2 \psi  -\psi = h \psi^* -i \gamma \psi,
\end{equation}
 was used to
 model
the nonlinear Faraday
resonance in a vertically oscillating water tank \cite{water,water_exp}
and the effect of phase-sensitive parametric
amplifiers on solitons in optical fibers \cite{optics}.
The same equation describes an easy-plane ferromagnet with a combination
of a static and  hf field in the easy plane
\cite{magnetism,BBK}, and  the planar weakly anisotropic
$XY$ model \cite{XY}.
It also serves as a continuum limit for
the parametrically driven
Frenkel-Kontorova chain (an array of diffusively
coupled pendula). The Frenkel-Kontorova system is regarded
as  a fairly realistic model of a number of physical and
biological systems and phenomena, including
  ladder
networks of discrete Josephson junctions,
charge-density-wave conductors,
 crystal dislocations in metals,
DNA dynamics
and proton conductivity in
 hydrogen-bonded chains \cite{Scott}.

The undamped undriven nonlinear Schr\"odinger equation
exhibits soliton solutions  which can travel with arbitrary
velocity and transport physical characteristics such as
mass, momentum and energy. (For the sake
of brevity,  we are making use of  the
hydrodynamical interpretation of the equation here.)
In equation (\ref{NLS}),
the second term in the right-hand side accounts for
dissipative
losses which occur in all physical systems.
 The dissipation
has two visible effects on the soliton:
it attenuates its speed and damps its amplitude.
The main purpose of the introduction of the  pumping
(represented by the first term in the right-hand side) is to compensate
for these losses.
The parametric forcing is well known to be capable of counterbalancing
the damping of the quiescent soliton's amplitude;
a natural question now is whether it can sustain its motion
with a nonzero velocity.

In fact the existence of travelling solitons is a nontrivial
matter even in the absence of damping.
The driving term $h \psi^*$ in Eq.(\ref{NLS})
breaks the galilean invariance of the
unperturbed nonlinear Schr\"odinger
 equation and hence one cannot obtain a
moving soliton simply by boosting a static one.
 However the  galilean or Lorentz symmetry is
not a prerequisite  for the existence of moving nonlinear waves.
For example,
 dissipative systems do not possess any symmetries of this kind
but are well known to support  stably propagating fronts
and pulses whose velocities are determined by parameters of the
model. In particular, travelling domain walls arise in the
parametrically-driven Ginsburg-Landau equations (where the motion is
due to  nongradient terms) \cite{Ginsburg_Landau}.
As far as solitons in Hamiltonian systems are concerned,
  the example
of dark solitons in the nonlinear Schr\"odinger equations
  suggests that they have
even  greater mobility than dissipative fronts and pulses.
Although in this case the galilean invariance is broken by
the presence of the nonvanishing background, the dark solitons
can propagate with  arbitrary speeds bounded only by the velocity of
sound waves \cite{dark_solitons,Bar,PAK}.

A number of nonstationary regimes were reported
in the water tank experiments, including
the formation of oscillating soliton
pairs \cite{water_exp}, but no steadily moving solitons were detected so far.
On the other hand,
 numerical simulations of the {\it undamped\/} parametrically
driven NLS
equation [Eq.(\ref{NLS}) with $\gamma=0$]
did exhibit travelling localised objects
\cite{Alex}.
It has remained an open question whether these moving objects preserve
their speed and amplitude, or attenuate and decay slowly due to the
emission of the second-harmonic radiation.
The aim of the present paper is to study
the
existence of steadily propagating solitons, and examine their
stability.
Here we are confining ourselves to the undamped situation
relegating the analysis of the effect of damping
to future publications.

In addition to their role in transport phenomena, stably moving solitons
are also of interest  as alternative attractors which may compete
with (static or oscillating) nonpropagating solutions.
We will demonstrate that stable travelling solitons do exist
in the parametrically driven nonlinear Schr\"odinger equation.
Moreover, there are parameter ranges where moving solitons
are stable whereas their quiescent counterparts are not.
Unstable  solitons are not meaningless either;
they arise as long-lived transients  and intermediate states
in spatio-temporal chaotic regimes. In this paper we will
 identify oscillatory and translational instabilities of
travelling solitons and simulate
their nonlinear evolution near the transition curves.

The structure of the paper is as follows. In section \ref{Prelimi}
we derive an {\it apriori\/} bound for the existence domain of
travelling solitons and introduce the linearised eigenvalue problem
for their stability analysis.
We also discuss some general properties of eigenvalues and eigenfunctions
and formulate a simple criterion for the onset of the nonoscillatory
instability: $\partial P/\partial V=0$, where $V$ is the velocity of
the steadily moving soliton, and $P$  the associated momentum.

In section
 \ref{explicit_for_V=0}
we present several explicit
quiescent ($V=0$) solutions, including a stationary complex
of two solitons,
and then derive the necessary condition for a static
solution to be continuable to nonzero velocities.
This condition is that the motionless solution should either
not have any ``free" parameters apart from the translational
shift, or, if there is an additional parameter $z$,
the equation  $\partial P/\partial z=0$ should be satisfied.
Here $P$ is the momentum of the motionless localised solution
(which, contrary to one's
mechanical intuition, is not necessarily equal to zero).
There are three static solutions satisfying the above condition,
two of which being the well-known constant-phase
 $\psi_+$ and $\psi_-$
solitons, respectively, while the third solution
looks like a pulse with a bell-shaped
modulus and twisted phase.

The most important results of this work are contained in
section \ref{Bif_diagram}
where we report on the numerical
continuation of various branches  of solutions
and their stability analysis. In agreement with the analytical
predictions of the preceding section, we find that each of the
above static solutions admits the continuation to nonzero $V$.
The stability properties of travelling solitons result from
 an intricate interplay of two types of instability, the
oscillatory and translational instability. In accordance with
the conclusions of section \ref{Prelimi},  the numerical analysis
of the linearised eigenvalues shows that the transition
curves of the translational
instability satisfy $\partial P/\partial V=0$. One interesting
conclusion of the stability analysis is that although
{\it quiescent\/} solitons are unstable for  driving strengths larger than
$h=0.064$,  there are stable {\it moving\/}
 solitons for any $0 \le h  \le 1$.
We discuss in detail the soliton's transformation as it is
continued in $V$, paying special attention to the dynamics of the
associated linearised eigenvalues on the complex plane.
Two different scenarios of the transformation are identified,
one occurring for small $h$ and the other one for larger
driving strength, and we also
describe an interesting cross-over from one to another.

Section \ref{simulations} is devoted to the direct numerical
simulations of the full time-dependent nonlinear Schr\"odinger
equation. We show that the evolution of both types of the soliton
instability leads, as $t \to \infty$, to the same asymptotic attractors.
 Finally,  section \ref{conclusions} contains
concluding remarks and outlines some open problems.

\section{Steadily  travelling waves: existence
and stability}
\label{Prelimi}
\subsection{Existence domain
and integrals of motion}
We will confine ourselves to localised travelling waves
of the simplest form,
$\psi(X,t)=\psi(X-Vt)$.
Transforming to the moving frame, these correspond
to time-independent soliton solutions of
the equation
\begin{equation}
\label{NLS_moving}
i \psi_t - iV \psi_x +\psi_{xx} +2 |\psi|^2 \psi  -\psi  = h \psi^*,
\end{equation}
where
 $x=X-Vt$.
We will search for these   static solutions by solving
 an ordinary differential
equation
\begin{equation}
\label{stationary_NLS}
-iV \psi_x +\psi_{xx} +2 |\psi|^2 \psi  -\psi
 = h \psi^*
\end{equation}
under the vanishing boundary conditions $|\psi(x)| \to 0$
as $|x| \to \infty$.
Here $h$ is always taken positive; negative $h$'s can be
recovered by the phase transformation $\psi \to i\psi$.

If $\psi(x)$ is a solution, then so is $-\psi(x)$.
Next, it is  straightforward to
notice that  if the function $\psi(x)$ describes
 a soliton travelling with the
velocity $V$, the function $ \psi^*$(x)  describes a soliton
moving with the velocity $-V$.
We will try to restrict ourselves to positive $V$'s
wherever possible;  we will only present negative
velocities where this may help visualising how different
branches of solutions are connected.
Since the soliton moving with
the velocity $-V$ is given by $\psi(-x)$, the above observation
tells us that either
$\psi^*(x)= \pm \psi(-x)$
(that is, one of
the real and imaginary part of the solution is even and the other one
odd), or there are {\it two\/} solutions associated with the same $V$.
(Here we are not making any difference
between  solutions which are  different just in the overal sign.)
In the latter case the solutions will not exhibit the
$\psi^*(x)= \pm \psi(-x)$  symmetry.

Next, it is easy to show that
 solitons cannot travel faster than a certain limit speed.
 Indeed, as $|x| \to \infty$, the
soliton's asymptotic tail decays as
 $\psi(x) \sim e^{- \kappa x}$, where
\begin{equation}
2 \kappa^2=
2-V^2
\pm
\sqrt{ \left(2-V^2 \right)^2 + 4(h^2-1)}.
\end{equation}
Large driving strengths, $h>1$, are of little interest to us as
in this case the zero background, $\psi(x)=0$,
is unstable with respect to continuous spectrum waves
\cite{BBK}.
Therefore we are not going to discuss this case here.
In the complementary region $h<1$, the complex structure of $\kappa$
depends on the value of the velocity.
When $V^2 <
 2- 2\sqrt{1-h^2}$,
there are four real exponents;
for
$ 2-2\sqrt{1-h^2}
<V^2 < c^2,$
where
\begin{equation}
 c = \sqrt{2 + 2 \sqrt{1-h^2}},
\label{V_bar}
\end{equation}
we have a quadruplet of complex $\kappa$'s,
and finally, for
$V^2>c^2$
all four exponents are imaginary.
Consequently,  there can be no exponentially localised solitons
travelling faster than $c$. Physically, $c$ represents
the minimum phase velocity of linear waves
governed by Eq.(\ref{NLS}), and our condition $V<c$ is essentially
an exclusion principle ruling out
  a resonance between
solitons and linear waves.

In the undamped case  the parametrically driven NLS equation (\ref{NLS_moving})
conserves the momentum,
\begin{equation}
P= \frac{i}{2} \int
\left(  \psi^*_x \psi -\psi_x  \psi^*
\right) dx,
\label{P}
\end{equation}
and energy:
\begin{equation}
E=
{\rm Re\/} \int \left( |\psi_x|^2 + |\psi|^2 - |\psi|^4 + h \psi^2
\right) dx.
\label{E}
\end{equation}
Noting that stationary solutions satisfy
\[
|\psi_x|^2 = |\psi|^2 - |\psi|^4 + h \, {\rm Re\/} \, \psi^2,
\]
Eq.(\ref{E}) can be rewritten as
\begin{equation}
E= 2 \,
{\rm Re\/} \int \left( |\psi|^2 - |\psi|^4 + h \psi^2
\right) dx.
\label{E2}
\end{equation}
Containing no derivatives,
this formula for energy of stationary solutions
has obvious advantages for the numerical  implementation.

\subsection{Linearised eigenvalue problem}

In this paper we solve the equation
(\ref{stationary_NLS}) numerically and examine the stability
of the resulting solutions by studying the associated
eigenvalue problem. This eigenvalue problem arises by
assuming a small perturbation of the form
\[
\delta \psi(x,t)= y(x)e^{\lambda t}, \quad
  y(x)= \delta u(x)+ i \delta v(x).
\]
Substituting into (\ref{NLS_moving})
 gives
\begin{equation}
{\cal H} Y = \lambda J Y,
\label{EV}
\end{equation}
where
the hermitean operator
${\cal H}$ has the form
\begin{eqnarray}
{\cal H} = I(-\partial_x^2 + 1)+ V J \partial_x+
\nonumber \\
+ \left(
\begin{array}{lr}
  h -6u^2 -2v^2 &  - 4uv \\
 -4uv & -h -6v^2-2u^2
\end{array}
\right);
\label{H}
\end{eqnarray}
 the matrix $J$ is given by
\begin{equation}
J=
 \left(
\begin{array}{lr}
0 & -1 \\ 1 & 0
\end{array} \right),
\end{equation}
and
the column-vector
  $Y(x)= ({\rm Re\/}\, y, {\rm Im\/}\, y)^T= (\delta u, \delta v)^T $.
In Eq.(\ref{H}) $I$ is the identity matrix,
and we have decomposed the stationary solution
as $\psi(x)=u(x) + iv(x)$.

For symmetric solutions satisfying $\psi^*(x)= \pm \psi(-x)$
eigenvalues will always come in ($\lambda, -\lambda$)-pairs.
This follows from the fact that for these solutions
changing $x \to -x$ in the operator (\ref{H}) amounts to changing
the sign of its off-diagonal elements, and hence if
$(\delta u(x), \delta v(x))^T$ is
an eigenfunction associated with an eigenvalue
$\lambda$, the column
$(\delta u(-x), -\delta v(-x))^T$
 will serve as an eigenfunction
associated with an eigenvalue $-\lambda$.
As far as the {\it zero\/} eigenvalue is concerned,
 it will have a twin
with the eigenfunction
$(\delta u(-x), -\delta v(-x))^T$ unless its eigenfunction
$y=\delta u + i \delta v$
satisfies the symmetry $y^*(x)= e^{i \varphi}y(-x)$,
where $\varphi=const$.

To complete the discussion of the spectrum structure, we mention
that there are two branches of the
continuous spectrum  lying on
the imaginary axis of $\lambda$: $\lambda= i \omega_{1,2}(k)$,
where
\[
\omega_{1,2}(k)= Vk \pm \sqrt{(k^2+1)^2-h^2},
\]
and $-\infty < k< \infty$. (We are still assuming $h<1$).
In the region $V^2<c^2$ which is of interest to us,
 the continuous spectrum has a gap:
$\omega_1(k)> \omega_0$, $\omega_2(k)<-\omega_0$,
where $\omega_0>0$. This gap  can harbour  discrete
  eigenvalues representing stable oscillation modes.

\subsection{Translational (nonoscillatory) instabilities}
\label{trans}
The aim of this subsection is to demonstrate
 that a pair of pure
imaginary eigenvalues
can collide at $\lambda=0$ and move
onto the real axis only at the point where
$\partial P/\partial V=0$.
This criterion is known in the context of dark
solitons of the undriven nonlinear
Schr\"odinger equations; see
\cite{Bar},\cite{PAK}. Here we simply adapt the proof given
in \cite{Bar} to the case of the equation with the parametric forcing.
An important assumption that we  make here,
is that the solution whose stability is being examined, does not
have any free parameters apart from the trivial translation parameter,
$x_0$.

First of all we need to make a remark on the integrable
case, $h=0$.
In this case  solutions of the ODE
(\ref{stationary_NLS})
 can be
obtained from a quiescent soliton
of Eq.(\ref{NLS}) by a Galilei transformation:
\begin{equation}
\psi(x)= e^{i (V/2) x} A \, {\rm sech\/} Ax,
\label{integrable}
\end{equation}
where $A=\sqrt{1-V^2/4}$.
For $h=0$ and any $|V|<2$, the  linearised operator ${\cal H}$ has
 four zero eigenvalues associated with two eigenvectors.
One of these eigenvectors originates from
 the translation symmetry and
the other one results from the phase invariance
of Eq.(\ref{stationary_NLS}).
The term $h \psi^*$ breaks the phase invariance and hence as
$h$ is increased from zero, one pair of eigenvalues $(\lambda,
-\lambda)$ moves away from the origin on the complex plane.
As $h$ and $V$ are further varied,  a pair of eigenvalues may
return to the origin. If the solution
of Eq.(\ref{stationary_NLS}) at the point of their return
is a member of a family parametrised by {\it two\/} free parameters, we will
have, again, four zero eigenvalues with two eigenfunctions.
(The eigenfunctions are simply derivatives of the solution with respect to
the free parameters.) Our analysis will not be applicable in this
case, and the equality $\partial P/\partial V=0$ does not have to be
valid at the return point.
(We will come across this type of a situation in section
\ref{further_ge_28} below.)
However, a more common situation is when
the solution at the return point is a member of a {\it one\/}-parameter
family. We will show that in this case the relation
$\partial P/\partial V=0$ does have to be in place.

Let  us  denote $V_c$
 the  velocity for which the
eigenvalue of the operator (\ref{EV})-(\ref{H}) vanishes.
We can develop the solution $\psi(V;x)$ in powers of  $\epsilon=V-V_c$:
\[
\psi(V;x)=\psi_0(x) + \epsilon \psi_1(x) +
\epsilon^2 \psi_2(x) +...,
\]
 where $\psi_0=\psi(V_c;x)$.
Accordingly, the operator
${\cal H}$ expands as ${\cal H}=
{\cal H}_0+\epsilon {\cal H}_1 +\epsilon^2 {\cal H}_2+...$.
If the eigenvalue $\lambda$
moves from  imaginary to the real axis, it is natural to
assume  that it admits an expansion of the form
\begin{equation}
\lambda= \epsilon^{1/2} \lambda_1
+ \epsilon^{3/2} \lambda_3 + \epsilon^{5/2} \lambda_5 + \dots .
\label{EV_expansion}
\end{equation}
 The associated eigenfunction is then developed as
\begin{equation}
Y(x)= Y_0(x) + \epsilon^{1/2} Y_1(x) + \epsilon Y_2(x)+... .
\end{equation}
When $\epsilon=0$, we have ${\cal H}_0 Y_0=0$, i.e.
$Y_0$ is a null eigenvector at the bifurcation point $V=V_c$.
Since we have assumed that $\psi(V_c,x)$ is a member of a
{\it one\/}-parameter family of solutions, the operator
${\cal H}_0$ has only {\it one\/} null eigenvector, and we have
to identify $Y_0= \Psi_0'(x)$.
Here
$\Psi_0$ is a column-vector formed by the
real and imaginary part of the soliton $\psi_0$:
$\Psi_0= (u_0,v_0)^T$. The prime indicates differentiation
with respect to $x$.

Next, setting the coefficient of $\epsilon^{1/2}$ to zero yields
\[
{\cal H}_0 Y_1= \lambda_1 J Y_0.
\]
Comparing this to the equation
\[
{\cal H}_0
  \left.
\frac{\partial \Psi}{\partial V}
\right|_{V=V_c}
=-J   \Psi_0',
\]
which arises from the differentiation of Eq.(\ref{stationary_NLS})
with respect to $V$, we get
\[
Y_1(x)=-\lambda_1
  \left.
\frac{\partial \Psi}{\partial V}
\right|_{V=V_c}.
\]
(In the above equations $\Psi=(u,v)^T$.)
The coefficient of $\epsilon^1$ produces
\[
{\cal H}_0 Y_2= \lambda_1 J Y_1 - {\cal H}_1 Y_0,
\]
which has bounded solutions if the right-hand side is orthogonal to the
null eigenvector of ${\cal H}_0$:
\begin{equation}
\lambda_1 \int Y_0 JY_1 dx-  \int Y_0 {\cal H}_1 Y_0 dx=0.
\label{100}
\end{equation}
The second term in  equation (\ref{100}) is readily shown to vanish ---
one only needs to expand the identity
${\cal H}   \Psi'=0$ in $\epsilon$.
(The coefficient of $\epsilon^1$ gives ${\cal H}_1
\Psi_0'=-{\cal H}_0  \Psi_1'$. Taking
the scalar product with $ \Psi_0'$ yields the
required $\int  \Psi_0' {\cal H}_1  \Psi_0' dx=0$.)
On the other hand, the first term in Eq.(\ref{100}) is
equal to $(\lambda_1^2/2) \partial P/ \partial V$.
Consequently, Eq.(\ref{100}) gives either $\partial P/\partial V=0$ or
$\lambda_1=0$. If we assume that $\lambda_1=0$, we will not be able to
conclude that $\partial P/\partial V=0$ at this order of the
expansion. However, the order $\epsilon^2$ will then give us
$\lambda_2^2 \partial P/\partial V=0$, which implies either
$\partial P/\partial V=0$ or $\lambda_2=0$. Proceeding by a
similar token we will eventually arrive
  at the equation
$\partial P/\partial V=0$ at some order  $\epsilon^n$ where $n$
is such that
$\lambda_n \neq 0$.
(Alternatively, we will have to conclude that all $\lambda_n=0$
and hence we are dealing with a symmetry eigenvalue which is equal to
zero for all $V$.)

Thus a pair of real or
pure imaginary eigenvalues of the same
magnitude and opposite sign,  can only collide for the value of $V$
 which satisfies $\partial P/ \partial V=0$.
Here we wish to re-emphasise that we have
obtained this conclusion  under the assumption that
 the geometric multiplicity of the zero eigenvalue is {\it not\/}
increased at the point of collision. A simple example when this
assumption is not valid,
 is furnished by the case $h=0$. In this case the momentum corresponding
to the soliton  (\ref{integrable})
is given by $P= V \sqrt{1-V^2/4}$. Although
$P$ has a maximum for $V=\sqrt{2}$,
 the stability properties of the undriven
soliton
do not change at this point.
The reason is that
for each
$V$ the operator ${\cal H}$ has {\it two\/} null eigenvectors
in this case,
and hence we cannot make the identification
 $Y_0= \Psi_0'$.
(Instead, $Y_0$ will be a linear combination of {\it two\/} zero modes.)
Consequently, the above proof becomes invalid in this case.

Finally, one can easily check that the above result does not
really depend on how the eigenvalue $\lambda$ expands in powers
of $\epsilon$. We assumed that the expansion (\ref{EV_expansion})
starts with terms of order $\epsilon^{1/2}$. This assumption
is natural and supported by the numerical evidence; however,
even if we had postulated  the expansion starting with
terms of order $\epsilon^{1/4}$, $\epsilon^{1/3}$
or say, $\epsilon$, we would have still arrived at the
{\it same\/}
 necessary condition for the zero crossing: $\partial P/\partial V=0$.

\section{Quiescent solutions ($V=0$)}
\label{explicit_for_V=0}
\subsection{The ``twist" soliton}

In order to continue  in $V$ it is useful to have  some
starting solutions for $V=0$. Two such stationary
 solutions are given by
\begin{mathletters}
\label{soliton}
\begin{eqnarray}
\psi_+(x)=A_+ {\rm sech\/} (A_+ x), \\
\psi_-(x)=iA_- {\rm sech\/} (A_- x),
\end{eqnarray}
\end{mathletters}
where
 $A_{\pm}^2= 1 \pm h$.
The soliton $\psi_-$ is unstable with respect to
a nonoscillatory mode for all $h$ \cite{BBK}.
The $\psi_+$ is stable for  $h< h_0=0.063596$
but developes  an oscillatory  instability as $h$ is
increased beyond $h_0$ \cite{BBK,Alex}.

 In this section we will
produce several more explicit solutions of the undamped,
parametrically driven NLS equation (\ref{NLS_moving}).
Writing $\psi=u+iv$, the stationary equation
(\ref{stationary_NLS}) transforms into the
system
\begin{eqnarray}
u_{xx} - u -hu + 2u(u^2+ v^2)=0, \label{system_1}
 \\
v_{xx} -v + hv +2v(u^2+v^2)=0.
\label{system_2}
\end{eqnarray}
We can try to find explicit solutions of this system by imposing some
plausible reductions,
for instance
by identifying $u^2+v^2$ with a function of $u$ and its derivatives: $u_x$,
$u_{xx}$ and so on.
In this case Eq.(\ref{system_1}) is an equation
for $u$ only, while
   Eq.(\ref{system_2}) becomes
a linear equation with variable coefficients.
 The simplest choice $u^2+v^2 = Cu^2$ leads to the
$\psi_+$ and $\psi_-$ solitons (\ref{soliton}). Another simple
possibility is to require that
\begin{equation}
u^2+v^2= Cu, \quad C={\rm const};
\label{reduction}
\end{equation}
this converts the first equation into the stationary KdV with the
well-known localised solution
\[
u=\frac34 \frac{1+h}{C}  {\rm sech\/}^2  \xi,
\quad
\xi=  \frac{\sqrt{1+h}}{2}x.
\]
The second equation has now the form of an eigenvalue problem
for the P\"oschl-Teller potential:
\begin{equation}
(-\partial^2_{\xi} +1  -6 {\rm sech\/}^2  \xi ) v
=E v,
\label{Poeschl}
\end{equation}
where $E=1+4(h-1)/(h+1)$. The  operator (\ref{Poeschl}) has two
localized eigenfunctions, $v_0=\alpha \, {\rm sech}^2 \xi$ associated with
an eigenvalue $E_0=-3$, and $v_1= \alpha \,
{\rm sech\/} \xi \, {\rm tanh\/} \xi$
with an eigenvalue $E_1=0$.
The first eigenfunction can satisfy the constraint
(\ref{reduction}) for no $\alpha$ while the second one will satisfy it
if we set $C^2=\alpha^2= \frac34 (1+h)$.

Noticing that $E=0$ corresponds to $h=\frac35$, we conclude that
 for $h=\frac35$ there is an explicit solution
of the form
\begin{eqnarray}
\label{twist}
u_T= \sqrt{\frac65} \, {\rm sech\/}^2 \xi;     \quad
v_T=  \pm \sqrt{\frac65} \, {\rm sech\/} \, \xi \, {\rm tanh\/} \, \xi,
\end{eqnarray}
where $\xi= \sqrt{\frac25} x$.
Similarly to the solitons $\psi_+$
and $\psi_-$, the modulus of the above solution is bell-shaped,
but, unlike the constant phase of the $\psi_+$
and $\psi_-$ solitons, the  phase of this solution varies.
 (In the case of the positive
sign in (\ref{twist}) the phase grows from $-\pi/2$ at $x=-\infty$
to $\pi/2$ at $x=\infty$.)
The solution looks like a pulse twisted by $180^{\circ}$ in the
$(u,v)$-plane.
  For this reason we will
be referring to Eq.(\ref{twist}) as the ``twist" soliton.

\subsection{The twist soliton as a bound state}
The system (\ref{system_1})-(\ref{system_2})
 appeared previously
 as a stationary system governing
light pulses in a birefringent optical fiber.
Using Hirota's approach, Tratnik and Sipe \cite{TS}
have obtained the following exact solution to
eqs.(\ref{stationary_NLS}),(\ref{system_1})-(\ref{system_2}):
\begin{mathletters}
\label{complex}
\begin{eqnarray}
\label{complex_0}
\psi= \psi(z;x)= u+iv, \\
\label{complex_1}
u= 2A_+ e^{\theta_2} D^{-1}
\left( 1 + e^{2 (\theta_1-\beta)}
\right), \\
\label{complex_2}
v= 2A_- e^{\theta_1} D^{-1}
\left( 1 - e^{2 (\theta_2-\beta)}
\right),
\end{eqnarray}
\end{mathletters}where

\begin{figure}
\begin{center}
\psfig{file=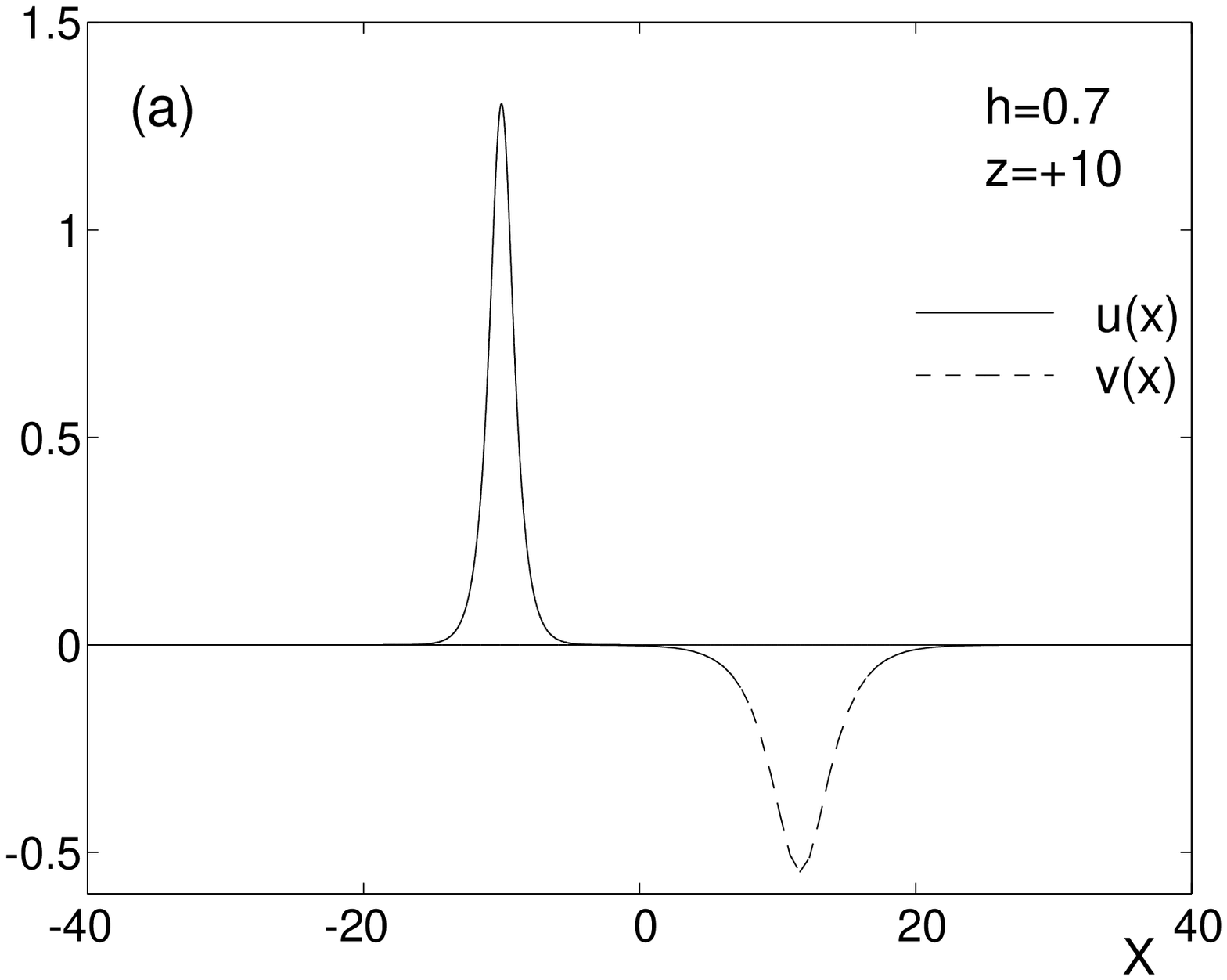,height=5cm,width=1.\linewidth}
\psfig{file=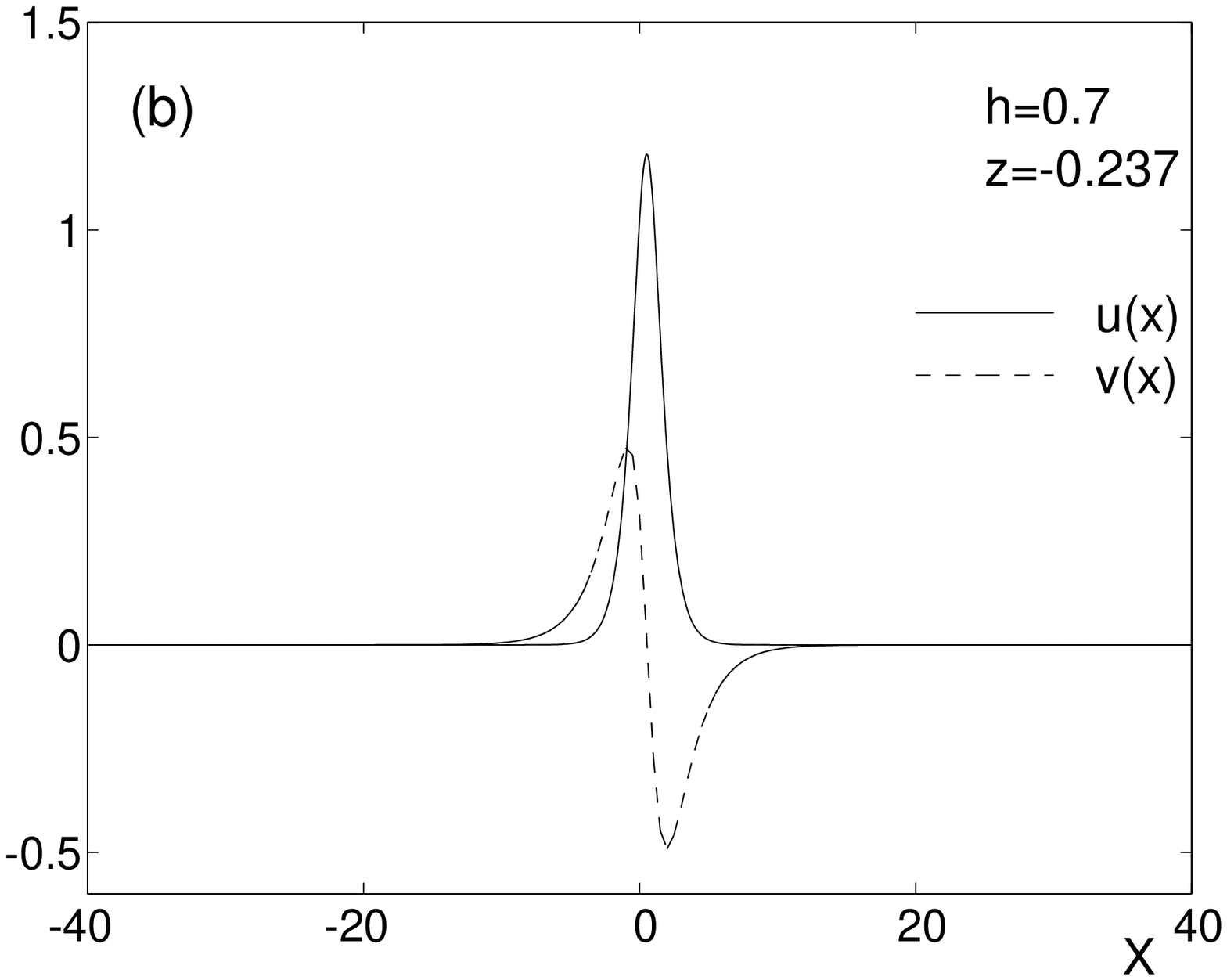,height=5cm,width=1.\linewidth}
\psfig{file=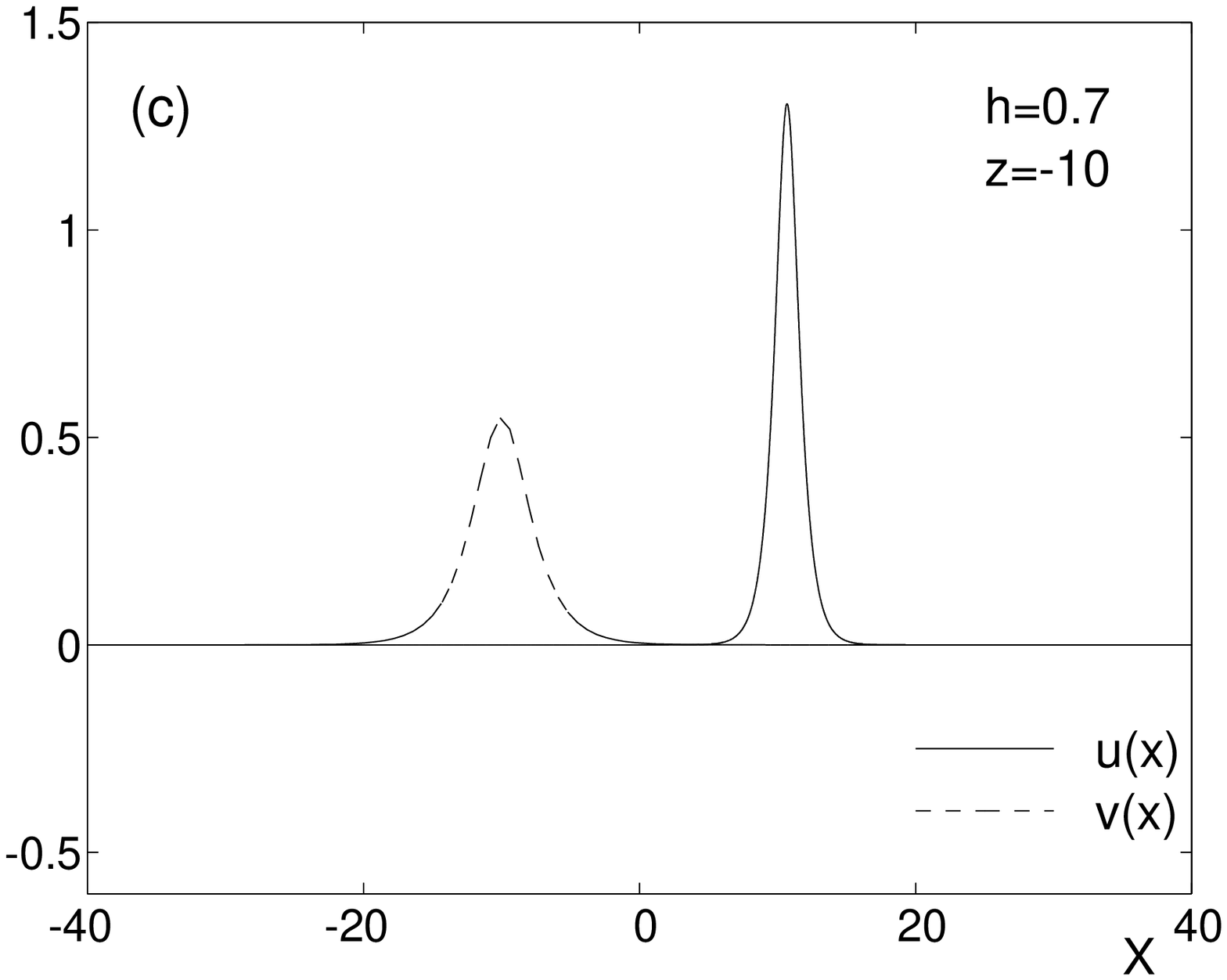,height=5cm,width=1.\linewidth}
\end{center}

\vspace{-20mm}
\label{two_solitons}
\end{figure}
{\sf {\bf Fig.1} Solution (\ref{complex})
for various $z$.
(a): $z=10$; (b): $z=\zeta$ with $\zeta$ as in (\ref{zeta}) (the twist
soliton); (c): $z=-10$.
Solid curve: real part; dashed line: imaginary part.}

\vspace{4mm}

\begin{eqnarray*}
D=1+e^{2 \theta_1} + e^{2 \theta_2} +
e^{2(\theta_1 + \theta_2 - 2\beta)}, \\
 \theta_1=A_-(x-z), \quad
\theta_2=A_+(x+z);
\end{eqnarray*}
the constant
\begin{equation}
\beta=\frac12 \ln \left( \frac{A_+ +A_-}{A_+ -A_-} \right) >0,
\label{beta}
\end{equation}
the amplitudes $A_{\pm}$ are as in (\ref{soliton}):
$A_{\pm} = \sqrt{1 \pm h}$,
and $z$ is a real parameter which can take arbitrary values.
The solution (\ref{complex})
 with $z=10$ and $-10$
is plotted in Fig.1, (a) and (c).
As is clear from the figure, for sufficiently
large $|z|$  the solution represents
a complex of two solitons, $\psi_+$ and $\psi_-$,
with the separation approximately equal to $2|z|$.
It is perhaps worth emphasizing here that the parameter $z$
has nothing to do with shifting  the solution as a whole,
$x \to x-x_0$, which is possible due to the translation
invariance of Eqs.(\ref{NLS})-(\ref{stationary_NLS}).
  (This overall shift parameter, $x_0$, is  disregarded
in  equations (\ref{twist})-(\ref{complex})
and in the remainder of this text).  The parameter $z$ is
nontrivial in the sense that the {\it shape\/} of the
solution depends on $z$.

It is straightforward to verify that our ``twist" solution is
a particular case of Eq.(\ref{complex}) with $h=\frac35$. Indeed,
choosing $z=-\frac18 \sqrt{\frac52} \, \ln 3$
in Eqs.(\ref{complex}),
we obtain
 the soliton (\ref{twist}) (the one with the
negative sign), centered at $x=-3z$.
Since the twist is a symmetric solution with {\it sech\/}-shaped
modulus,
it would be
difficult to interpret it as a bound state of the $\psi_+$
and $\psi_-$ without embedding it into a
broader family of solitonic complexes.
In fact, a similar symmetric solution exists for any $h$,
not only for $h = \frac35$.
To see this, we notice a simple relation between two solutions
of the form (\ref{complex}) --- one with the parameter value
$z=\zeta+ \xi$ and the other one with $z=\zeta-\xi$:
\begin{equation}
\psi(\zeta + \xi; \eta -y)= \psi^*(\zeta-\xi; \eta+y).
\label{symsym}
\end{equation}
Here $\zeta$ and $\eta$ are defined by the driving strength $h$:
\begin{equation}
\zeta= -\frac{\beta}{2} \left( \frac{1}{A_-}- \frac{1}{A_+} \right)<0
\label{zeta}
\end{equation}
and
\begin{equation}
\eta= \frac{\beta}{2} \left( \frac{1}{A_-} + \frac{1}{A_+} \right),
\end{equation}
while $\xi$ and $y$ can take arbitrary values.
The relation (\ref{symsym}) implies that the
solution (\ref{complex}) with $z=\zeta$ is symmetric about the point
$x=\eta$:
\begin{equation}
\psi(\zeta ; \eta -y)= \psi^*(\zeta ; \eta+y).
\label{symmetty}
\end{equation}
That is, the real part of this solution is even and imaginary
part odd with respect to  $x=\eta$:
\[
u(\eta-y)=u(\eta+y), \quad v(\eta-y)=-v(\eta+y).
\]
(See Fig.1 (b).) This particular representative of the
family (\ref{complex})
 will play a special role in what follows.
For the ease of reference we are retaining the  name
``twist" for this symmetric solution --- for all $h$.

\subsection{The moving soliton bifurcation}
\label{moving_bifurcation}
Suppose the equation
(\ref{stationary_NLS}) has a
one-parameter family of quiescent solutions $\psi(z;x)$.
Here $z$ can be
any nontrivial parameter;
the only requirement is that  $z$ should
not be just an overall shift in $x$.
One such family is given by Eq.(\ref{complex}) and
 there can also be other families for which $\psi$ is not
available explicitly.
We will show in this section that in order for
a solution  with some $z=z_0$ to be continuable to
nonzero $V$, the corresponding momentum integral
should satisfy
\begin{equation}
\left.
\frac{\partial P}{\partial z}
\right|_{z=z_0} =0.
\end{equation}

Let us assume that
 Eq.(\ref{stationary_NLS}) with $V \ne 0$
has a solution $\psi(x)$, and that this solution is an analytic function
of $V$
in some neighbourhood of $V=0$. Then we can expand it in the Taylor
series
\begin{equation}
\psi(x)=\psi_0(x) + V \psi_1(x) + V^2 \psi_2(x) +...,
\label{Taylor}
\end{equation}
where $\psi_0(x)= \psi(z_0;x) = u_0+iv_0$ is some
representative of the family of
``motionless" solutions $\psi(z;x)$ with the parameter value $z_0$.
Substituting (\ref{Taylor}) into (\ref{stationary_NLS})
and equating coefficients of like powers of $V$, we get,
at the order  $V^1$:
\begin{equation}
{\cal H} \left(
\begin{array}{c}
u_1 \\ v_1
\end{array}
\right)
= J \partial_x
 \left(
\begin{array}{c}
u_0 \\ v_0
\end{array}
\right).
\label{first_order}
\end{equation}
Here $u_1+iv_1=\psi_1$ and the operator ${\cal H}$ is given by Eq.(\ref{H}).
The equation (\ref{first_order}) is solvable
in the class of square integrable functions if
the vector in the right-hand side is orthogonal to
all homogeneous solutions, i.e. to all null eigenvectors of the
operator ${\cal H}$. Since there is a family of
``motionless" solutions parametrized by $z$
and by an arbitrary spatial shift $x_0$
(which we have disregarded so far), the operator
${\cal H}$ has two zero modes. One is the translation
mode $\partial_x \psi_0= \partial_x( u_0+ i v_0)$;
the corresponding solvability condition is trivially satisfied:
\[
\int \partial_x (u_0, v_0)
J
 \partial_x
\left(
\begin{array}{c}
u_0 \\ v_0
\end{array}
\right) dx =0.
\]
The other zero mode is given by the derivative
$\partial_z \psi_0= \partial_z u_0 + i \partial_z v_0$.
The associated solvability condition reads
\[
0=\int (\partial_z u_0, \partial_z v_0)
J
 \partial_x
\left(
\begin{array}{c}
u_0 \\ v_0
\end{array}
\right) dx= -\frac12
\frac{\partial P}{\partial z},
\]
where $P$ is the momentum integral (\ref{P}). Consequently,
a solution with nonzero $V$ can only detach from
the $V=0$ branch at the point where $\partial P/\partial z=0$.

Coming back to our explicit solutions, the $\psi_+$
and $\psi_-$ solitons do not have any free parameters apart from
the trivial position shift. Consequently, both solutions
are continuable to nonzero $V$. Next, we
have a family of solitonic complexes
(\ref{complex}) with a nontrivial parameter $z$.
As one can easily check, the
momentum of the complex (\ref{complex}) as a function
of $z$ has a single
minimum for some finite $z=z_0$ and  tends to zero as
$z \to \pm \infty$.
To find $z_0$, we notice that the relation
(\ref{symsym}) implies
\[
P(\zeta + \xi)=P(\zeta- \xi).
\]
This means that the function $P(z)$ is even with respect to
the point $z=\zeta$ and therefore,  $\zeta$ is the point of the minimum:
$z_0=\zeta$.
Thus, the
 only representative of the family of the two-soliton
complexes (\ref{complex}) that can be continued to
nonzero $V$, is our twist soliton, $\psi(\zeta;x)$.
(To be more precise, there are {\it two\/} twist solutions,
one with positive and the other one with negative momentum.
This is related to the fact that when $V=0$, we can generate
new solutions to the system (\ref{system_1})-(\ref{system_2})
by changing the sign of just one component $u$ or $v$.)

\section{Bifurcation diagram}
\label{Bif_diagram}
We used a predictor-corrector continuation
algorithm with a fourth-order Newtonian solver
 to continue
solutions of
equation (\ref{stationary_NLS}) in $V$.
Since  derivatives of the
  momentum integral (\ref{P}) determine stability and branching
properties of solutions, the momentum was our natural choice
for the  bifurcation measure.
Eq.(\ref{stationary_NLS})
was solved under the vanishing boundary
conditions $\psi \left( \pm L/2
\right)=0$. We used   $L=200$
(except the cases where we had to extend the interval
to account for slow decay of solutions)
and the discretisation stepsize $\Delta x=0.005$.
The eigenvalue problem (\ref{EV}) was solved on the interval
$(-50,50)$.
Here we utilised the Fourier
method, typically with 600 harmonics.

\subsection{The travelling  $\psi_-$ soliton}

We start our description with
the branch  departing from the quiescent soliton
$\psi_-$. For every
 $h$ this branch continues all way to
$V=c$,
where $c$ is the minimum phase velocity
of linear waves given by Eq.(\ref{V_bar}).
(See Fig.2).  As $V \to c$,
the decay rate of $\psi(x)$  decreases and the soliton merges with
the zero solution, with the momentum $P$ tending to zero.
We plotted $P(V)$ for $h=0.1, 0.3, 0.5, 0.7$ and $0.9$ in Fig.2.
For technical reasons
we could not connect the curve
$P(V)$ to zero although we were able to approach
the value $V=c$ as close as the fourth digit after the decimal point.
 (The problem is that since the decay rate of the solution
decreases, one has to increase the length of the integration interval
--- and this cannot be done indefinitely.) The only curve which is
connected to zero in Fig.2, is the one for the undriven case
$h=0$. In this case we enjoy an explicit solution (\ref{integrable})
with the momentum $P=V \sqrt{1-V^2/4}$.

\end{multicols}
\begin{figure}
\begin{center}
\psfig{file=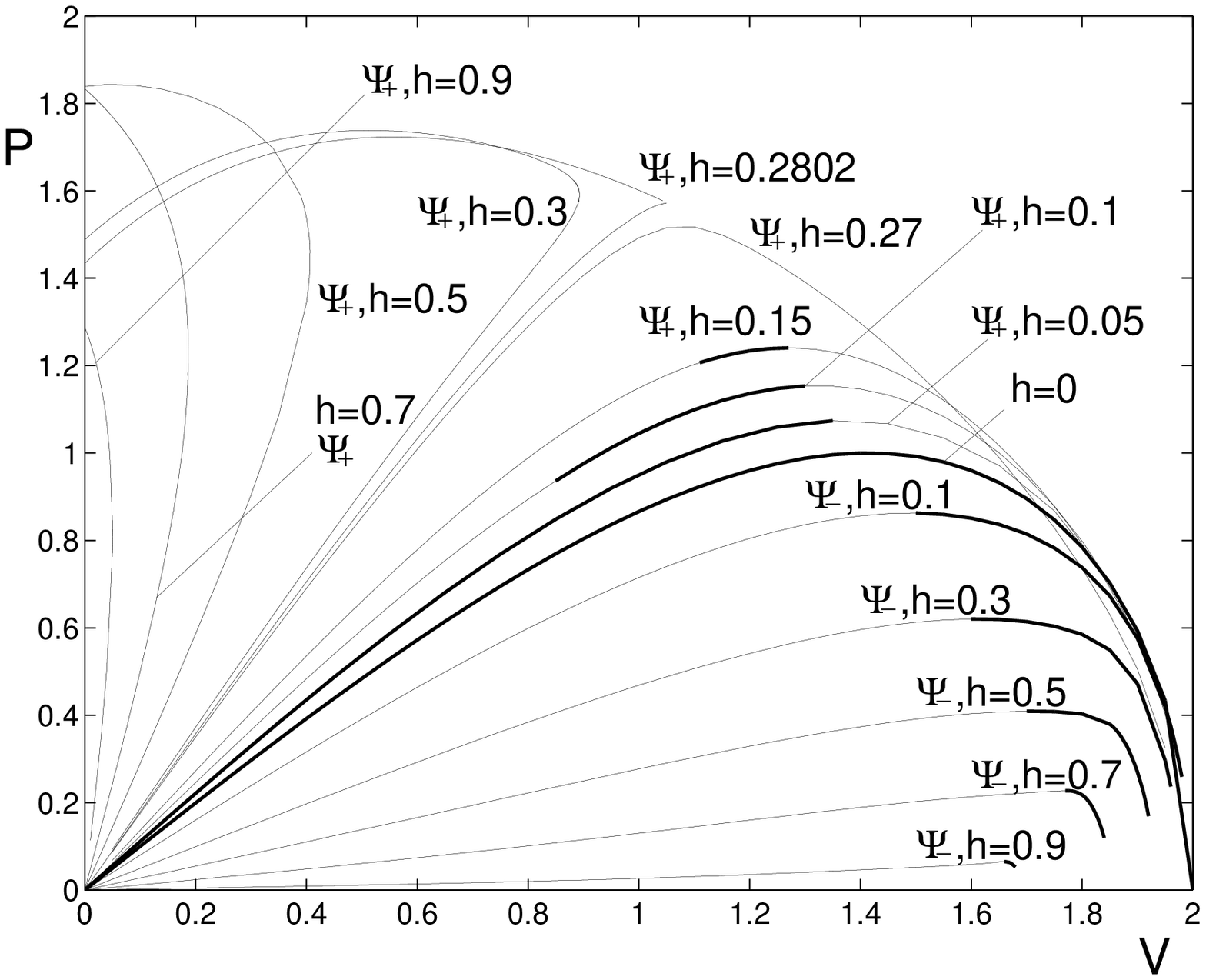,height=10cm,width=1.\linewidth}
\end{center}

\vspace{-12mm}
\label{P(V)}
\end{figure}
{\sf {\bf Fig.2} The  momentum
of the  $\psi_+$ and $\psi_-$ solitons as a function of their
velocities.
Thick respectively thin lines depict stable respectively unstable
solutions. Note that  as $h \to 0$,
 the stability domain of the $\psi_-$
tends to $\sqrt{2} \le V \le 2$ and that of the $\psi_+$
to $0 \le V \le \sqrt{2}$. The whole of the $h=0$ branch is stable.
}
\vspace{3mm}
\begin{multicols}{2}

 For each $h$ the momentum of the soliton  has a
single maximum on this branch, at $V=V_c$
(Fig.2).
To the left of $V_c$ the linearised operator (\ref{EV})-(\ref{H})
has a pair of real eigenvalues $ \pm \lambda$ and consequently,
the soliton $\psi_-$, which is well known to be unstable
for $V=0$ \cite{BBK}, remains unstable for small nonzero
velocities.
As $V$ approaches $V_c$,
 the two eigenvalues  converge at the origin
on the complex plane,
 with
the associated eigenfunctions
tending to the translation mode
$\Psi_0'(x)$.
Increasing $V$  past $V_c$, the eigenvalues
move onto the imaginary axis and hence the
$\psi_-$ soliton becomes stable for sufficiently large
velocities (where $\partial P/\partial V<0$).
This change of stability properties
  is in exact agreement with the scenario described in
section \ref{trans}.
\subsection{The travelling  $\psi_+$ soliton; $h \le 0.25$}
In agreement with  conclusions of section \ref{moving_bifurcation},
we have found that
the soliton $\psi_+$ is also continuable to nonzero
velocities.
Unlike the $\psi_-$-branch where the final product of the continuation
process
does not depend on the value of $h$, the transformation
scenario
  of the soliton $\psi_+$
is different for different $h$.
For $h< 0.28$
the fate of
the soliton $\psi_+$ is  similar
to that of the $\psi_-$. As $V \to c$,
the soliton develops oscillations
on its tails; the width
of the resulting oscillatory ``wavepacket" grows
and the amplitude decreases,
until the solution becomes equal to zero everywhere.
The momentum  $P(V)$ tends to zero as
 $V \to c$
and has a single  maximum
at some $V=V_c$.
Stability properties of the $\psi_+$ soliton depend on whether
  $h$ is smaller than
 0.064, lies between  0.064 and 0.25, or is
greater than 0.25.

Let, first, $h \le 0.25$. As $V$ is increased to $V_c$,
 a pair of  imaginary eigenvalues $\pm \lambda$ collides at
the origin on the complex plane
and moves onto the real axis. This is in contrast to the
$\psi_-$ soliton, where two {\it real\/} eigenvalues
converged at the origin as $V$ was increased.
For $V>V_c$
(i.e. where the slope $\partial P/\partial V$ is {\it negative\/}),
 the $\psi_+$
 soliton becomes  unstable. (Note that the $\psi_-$ was unstable
for {\it positive\/} slopes $\partial P/ \partial V$.)
As in the
$\psi_-$ case, this is the translational, or nonoscillatory, instability.
Since, as we checked, the eigenfunctions associated with
the colliding imaginary eigenvalues tend to
$\Psi_0'(x)$ as $V \to V_c$, this instability is of the type
analysed in section \ref{trans}. The numerically detected
instability boundary, defined  by the condition $\partial P/\partial V=0$,
is in exact correspondence with our analytical predictions.

In addition, for some $V$
 the $\psi_+$ soliton exhibits the oscillatory instability.
(The oscillatory instability sets in when
four  eigenvalues collide, pairwise, on the imaginary axis and emerge
into the complex plane.)
For $h \le 0.064$ the oscillatory instability does not
arise for any $V$ and hence the whole range $0<V<V_c$ is stable.
(See the $h=0.05$ curve in Fig.2).
For  $0.064 <h \le 0.25$ the
oscillatory instability occurs for  $V$
zero and small,
but as $V$ is increased,
 both imaginary and real
parts of the ``unstable" eigenvalues decay, with the real parts
 decaying faster.
Eventually the eigenvalues $\pm \lambda, \pm \lambda^*$ converge,
pairwise, on the imaginary axis and
 the soliton stabilizes.
Increasing $V$ still further, two of the resulting imaginary eigenvalues,
$\lambda$ and $-\lambda$,
start approaching each other.
At $V=V_c$ where $\partial P/\partial V=0$,
they collide and move onto the real axis.
 The soliton looses its stability once again --- this time
to a nonoscillatory mode.
This scenario is exemplified by the curves $h=0.1$
and $h=0.15$ in Fig.2.

An interesting phenomenon occurs in the undriven case,
$h=0$. As we mentioned,
in this case the stationary solution of Eq.(\ref{NLS_moving})
 is available in explicit form, Eq.(\ref{integrable}).
We also noted that the associated momentum $P=V \sqrt{1-V^2/4}$
has a maximum for finite $V=\sqrt{2}$.
  On the other hand, in any small-$h$ neighbourhood
of this solution there are solitons $\psi_+$ and $\psi_-$,
 of which $\psi_+$ is stable only for $\partial P/\partial V>0$
(i.e. for $V< \sqrt{2}$) and $\psi_-$ is stable only
for $\partial P/\partial V<0$ (i.e. for $V> \sqrt{2}$).
How can this be reconciled with the fact that the
 solution (\ref{integrable}) is  stable
 for all $V$ for which it exists
($V \le 2$)?

The answer is related to the
behaviour of the zero eigenvalue associated with the
phase invariance of Eq.(\ref{NLS_moving}) with $h=0$.
As $h$ increases from zero, this eigenvalue moves away
from the origin. On the $\psi_+$ branch, it
passes  onto the imaginary axis, while on the $\psi_-$ branch,
the eigenvalue moves onto the real axis instead.

\subsection{The travelling  $\psi_+$ soliton; $h > 0.25$}

Let now $0.25 < h <0.28$, and assume we are moving along
the $\psi_+$ branch in the direction of larger $V$.
 For small
$V$ we have a quadruplet of complex eigenvalues
$\pm \lambda$, $\pm \lambda^*$ implying the oscillatory
instability. As $V$ is increased, both imaginary and real
parts decay --- as in the $h \le 0.25$
case. However, this time  imaginary parts decay faster
than the real parts,
and the two pairs of eigenvalues converge on the {\it real\/}
axis.  For  velocities above this point the
 oscillatory instability is replaced by
the translational instability.
As $V$ is increased further, one pair of the newly
born real eigenvalues grows in absolute value whereas the
other pair decreases in magnitude.
At the point $V=V_c$ where $P(V)$ reaches its maximum,
the latter pair converges at the origin and moves onto the
imaginary axis. (This does not render the soliton stable
though, as the other pair remains on the real axis.)
This scenario is exemplified by the curve $h=0.27$ in Fig.2.

Next, let  $h$ be greater than $0.28$. For these $h$
the branch $P(V)$ emanating from the origin,
 turns back at some  $V=V_{\rm max}$ (Fig.2),
with the derivative $\partial P/ \partial V$ remaining
strictly positive for all $V \le V_{\rm max}$.
Below we will describe
the transformation this solution undergoes when
 continued
beyond the ``turning point", while
 here we only wish to
emphasize that no new zero eigenvalues can appear at this
 point.
The reason is that $V=V_{\rm max}$ is a bifurcation point
of solutions of the {\it ordinary\/} differential equation
(\ref{stationary_NLS}) but not of the
{\it partial\/} differential equation (\ref{NLS})-(\ref{NLS_moving}).
(In other words, $V$ is an ``internal" parameter
characterising the solution and not
an  ``external" control parameter.)
Indeed, the soliton is a member of a two-parameter
($x_0$ and $V$)
family of solutions of
Eq.(\ref{NLS})-(\ref{NLS_moving})
and hence
for {\it any\/}
$V$
 there are two zero eigenvalues
in the spectrum of the linearised operator ${\cal H}$.
Consequently, although being  a turning point for
the ODE (\ref{stationary_NLS}), the value $V=V_{\rm max}$
 is in no way special as far as the
 PDE (\ref{NLS})-(\ref{NLS_moving})
and its  linearisation are concerned.
No changes of the soliton's
stability properties occur at this velocity.

How does one type of behaviour of the curve $P(V)$,
occurring for $h > 0.28$, replace the other one,
arising for $h < 0.28$? We  scanned the
interval   $0.28000 < h< 0.28020$ and discovered
a tiny region of transitional behavior, around $h=0.28005$.
For this $h$,  $P(V)$ grows until it reaches a maximum
at $V_c=1.051$ and then starts decreasing, as in the
case of $h < 0.28$. However, the curve does not decay all
way to $P=0$ as would be the case for $h < 0.28$,
but reaches a minimum at $V_{cc}=1.0563$. After
that, the momentum  starts growing, and, at $V_{\rm max}= 1.0565$,
the curve $P(V)$ turns back --- as for $h > 0.28$!
To give an idea of how small this window of transitional
behaviour is, it suffices to say that for
$h= 0.28000$  the momentum  $P(V)$ tends to 0 as $V \to c$,
whereas for $h$ as close as
$0.28010$, the curve $P(V)$ has a ``turning point".

\begin{figure}
\begin{center}
\psfig{file=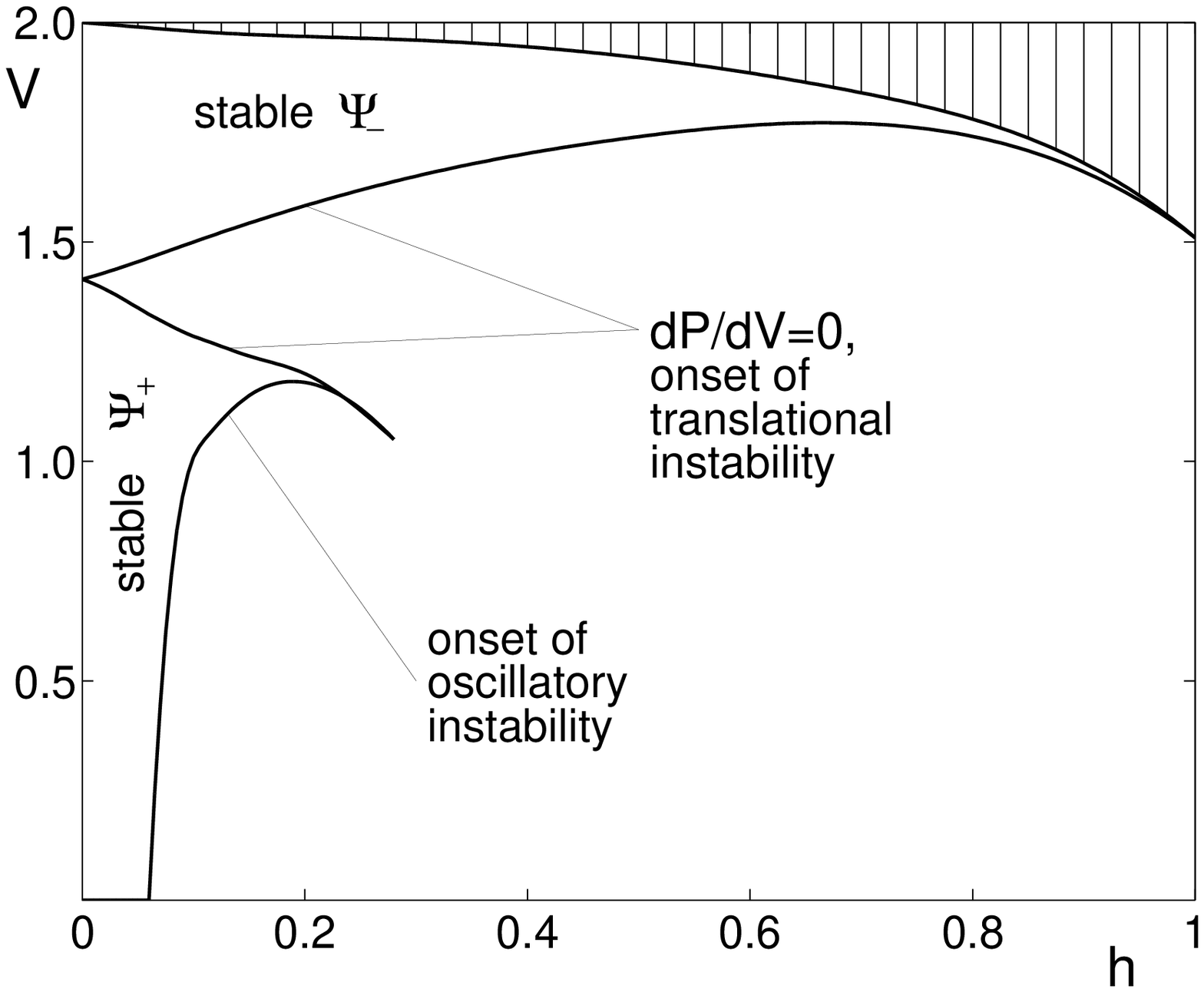,height=5cm,width=1.\linewidth}
\end{center}

\vspace{-12mm}
\label{stab_diagram}
\end{figure}
{\sf {\bf Fig.3}  Stability diagram for the $\psi_+$
and $\psi_-$ solitons on the $(h,V)$-plane.
In the region marked ``stable" one of the two one-soliton
solutions is stable whereas the other one is not.
Across the solid line, the corresponding
 soliton looses its stability
to an oscillatory or monotonically growing mode.
No solitons exist in the dashed region.}
\vspace{5mm}

What happens to the $\psi_+$ soliton with $h>0.28$
(more precisely, with $h \ge 0.28010$)
as we continue it beyond the turning point?
 Fig.4 (a) shows the momentum as
a function of $V$. The point of intersection with the
vertical axis $V=0$ corresponds to the twist solution
(Eq.(\ref{complex}) with $z=\zeta$ given by
 Eq.(\ref{zeta}).)
On the diagram Fig.4, it is marked as $\psi_T$.
As we mentioned before, the $(V=0)$-twist is a representative
of a two-parameter family of stationary
solutions of Eq.(\ref{NLS_moving}). Consequently, there
should be four zero eigenvalues in the
spectrum of the operator ${\cal H}$ in
this case, with two linearly independent eigenfunctions given by
$ \left. \partial_x \Psi(z;x) \right|_{z=\zeta}$
and $ \left. \partial_z \Psi(z;x) \right|_{z=\zeta}$.
(Here $\Psi(z;x)$ is a two-component vector formed by the
real and imaginary parts of Eq.(\ref{complex}).)
Numerically, we observed that
as we approach the ($V=0$)-twist from the direction of
positive $V$,
a pair of opposite eigenvalues
converges at the origin on the complex plane.
 The curve $P(V)$ does not
have an extremum at this point and this may seem to be in
contradiction with  predictions of section \ref{trans}.
The paradox is resolved as soon as one recalls that
the extremality condition  $\partial P/\partial V=0$ was derived
under the assumption that there is only {\it one\/} eigenvector
associated with the zero eigenvalue whereas we have {\it two\/}
linearly independent null eigenvectors
in the case at hand.

As we continue  further into the region $V<0$,
the twist gives rise to a variety
of multisoliton
complexes; we shall describe them in the next subsection.
Here we will restrict ourselves to the region $V>0$ where
this branch can still be regarded as a branch of one-soliton
solutions.
Although these solutions  undergo  similar transformations
for all $h$ in the interval $(0.28, 1)$, there are
a few differences with regard to  the  trajectories of eigenvalues
on the complex plane.
One difference worth to be mentioned is that
 for the driving strength $h =  0.3$
and larger $h$,   the  quadruplet of complex
$\lambda$ persists
on the entire upper branch of $P(V)$ (i.e. for all $V>0$).
This is in contrast to the case of $ 0.25< h < 0.28$, where
the complex quadruplet converges on the  real axis.
Near the left end of the interval $0.28 <h <1$
(e.g. for $h=0.2802$), we have an intermediate pattern.
Similarly to the case $h<0.28$, here the complex quadruplet
converges on
the real axis somewhere on the lower branch of $P(V)$
(i.e. before the turning point), but
 as we  move onto the upper branch, the two emerging real
pairs reunite quickly
and the complex quadruplet reappears.

Next, as we know, there are only two points where
 a pair of eigenvalues
 can pass from the real onto the imaginary axis, or vice versa.
One point is $V=V_c$ where $\partial P/\partial V=0$,
and the other one is $V=0$.
Therefore the dynamics of eigenvalues
 depends on which of the two points comes first,
or, equivalently,
whether the upper branch
of $P(V)$ has the maximum for positive or negative $V$.
For smaller values of $h$ in the interval $(0.28,1)$
(e.g. $h=0.3$), where $V_c>0$,
 two
 imaginary eigenvalues move to the real axis at $V=V_c$.
These imaginary eigenvalues have detached from the continuous
spectrum somewhere before the turning point
(i.e. on the lower branch of $P(V)$.)
The two
newly born real eigenvalues first diverge from the origin but then
reverse and, at $V=0$, move back onto the imaginary axis.
For larger $h$ (e.g. $h=0.7$), where
 $V_c<0$, the pattern is different.
 Here, the two imaginary eigenvalues become real
not at the point $V_c$  but at $V=0$. Subsequently, as we continue the
branch to negative velocities, another pair of imaginary
eigenvalues detaches from the continuum and
at the point $V_c<0$ two (imaginary or real) eigenvalues
pass through the origin.

Fig.3 shows the  stability diagram of the
$\psi_+$ and $\psi_-$ solitons on the $(h,V)$-plane.
For the $\psi_+$ soliton, the  range of stable velocities
approaches $0 \le V <\sqrt{2}$
as $h \to 0$,
while the stability range of $\psi_-$ tends to
$\sqrt{2} < V \le 2$. Finally, the domain
of stability in the $h=0$
case is the union of the above two ranges: $0 \le V \le 2$.

\subsection{Other branches; $h>0.28$}
\label{further_ge_28}

As we continue it
 to negative velocities, the twist
(we are using this name here for $V \neq 0$-deformations of the
quiescent twist solution)
 gradually
transforms into a complex of two twists
(plotted in Fig.5(a)).
A further continuation of this branch takes us,
via several ``turning points",
to a solution that can be interpreted as
 an
association of the twist and
two $\psi_-$ solitons of  opposite polarities
(denoted $\psi_{(-T-)}$). This solution is
depicted in Fig.5 (b).

\begin{figure}
\begin{center}
\psfig{file=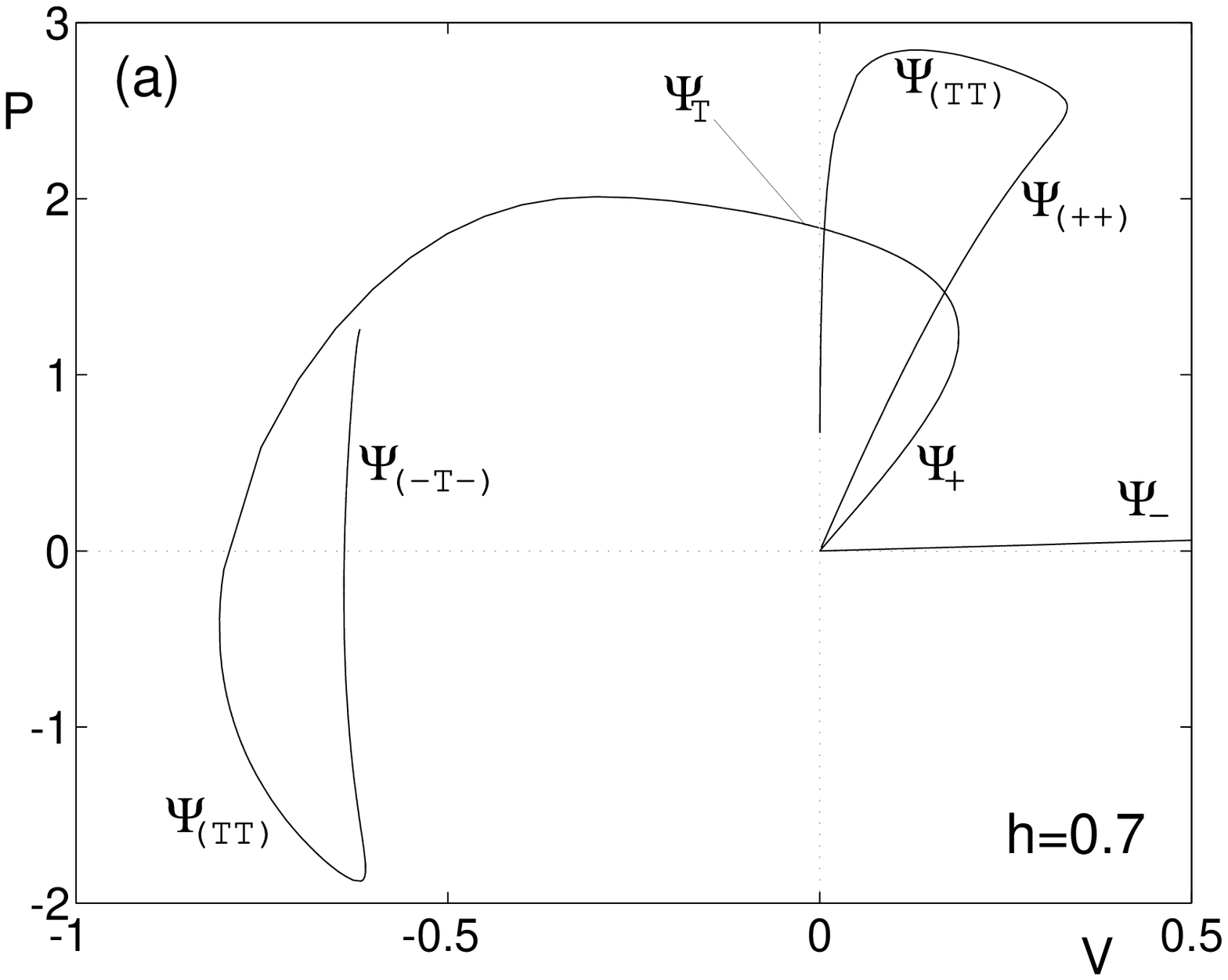,height=5cm,width=1.\linewidth}
\psfig{file=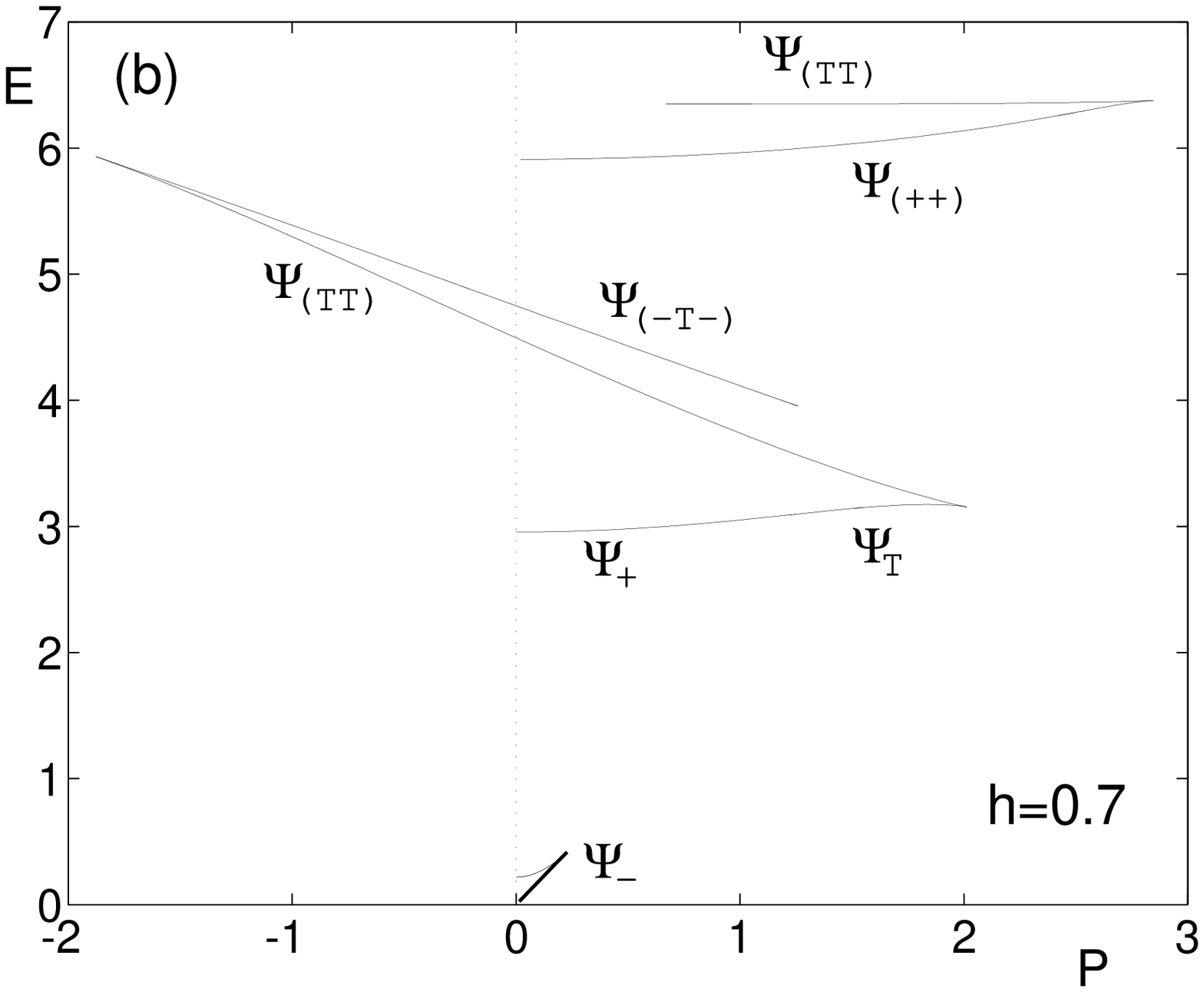,height=5cm,width=1.\linewidth}
\end{center}

\vspace{-12mm}
\label{bif_diagram}
\end{figure}
{\sf {\bf Fig.4} (a): The full bifurcation diagram for $h=0.7$.
All branches shown in this figure are unstable. (For
$h=0.7$ the soliton
$\psi_-$ is stable at large velocities
but the stable region lies beyond
the frame of this figure --- see Fig.2.)
More solution branches can be obtained by the
reflection $V \to -V$, $P \to -P$.
(b): The corresponding $E(P)$ dependence.
The stable branch of $\psi_-$ is depicted by a thick line
at the bottom of the figure.
 }
\vspace{5mm}

Fig.4 (b) shows the energy of different branches
as calculated by Eq.(\ref{E2}).  We have eliminated the
dependence on the soliton's velocity between $P(V)$
and $E(V)$ to obtain $E$ as a function of $P$.
The purpose of this ``Legendre transformation" is the
following. The stationary equation (\ref{stationary_NLS})
can be regarded as a condition that the energy
(\ref{E}) be stationary under the fixed momentum (\ref{P}):
$(\delta E)_P=0$ or, equivalently, $\delta (E-VP)=0$,
where $V$ is the Lagrange multiplier.
Consequently, a steadily travelling soliton satisfies the
relation of the Hamilton mechanics, $V=\partial E/\partial P$,
and its velocity is given by the slope of the corresponding $E(P)$
branch in Fig.4 (b). The relation $\delta E=V \delta P$
implies that the curves $E(V)$ and $P(V)$  have extrema
at the same set of points $V=V_c$
which will appear as cusps  on the $E(P)$ plot.
Since the  function $P(V)$ has extrema precisely at points where linearised
eigenvalues move from the imaginary to real axis,
 branches of the $E(P)$ curve separated by the cusps may have
different stability properties. One  such  change
of stabilities does indeed
occur in Fig.4 (b) where the $\psi_-$-curve is seen to be
partitioned into stable and unstable branch.
Solitons
on the stable branch have lower energies than unstable solitons
with  the same values
of the momentum.

\begin{figure}
\begin{center}
\psfig{file=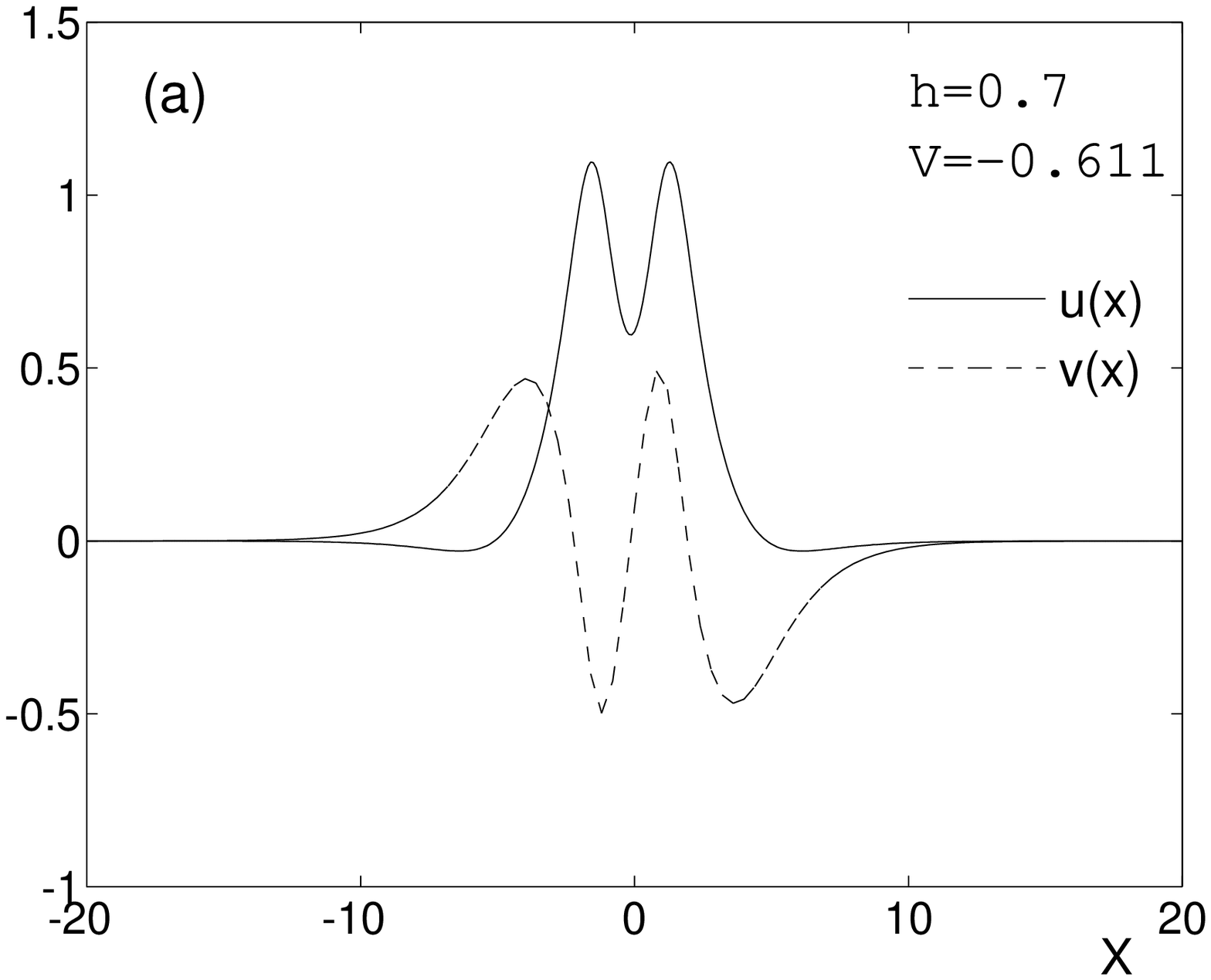,height=5cm,width=1.\linewidth}
\psfig{file=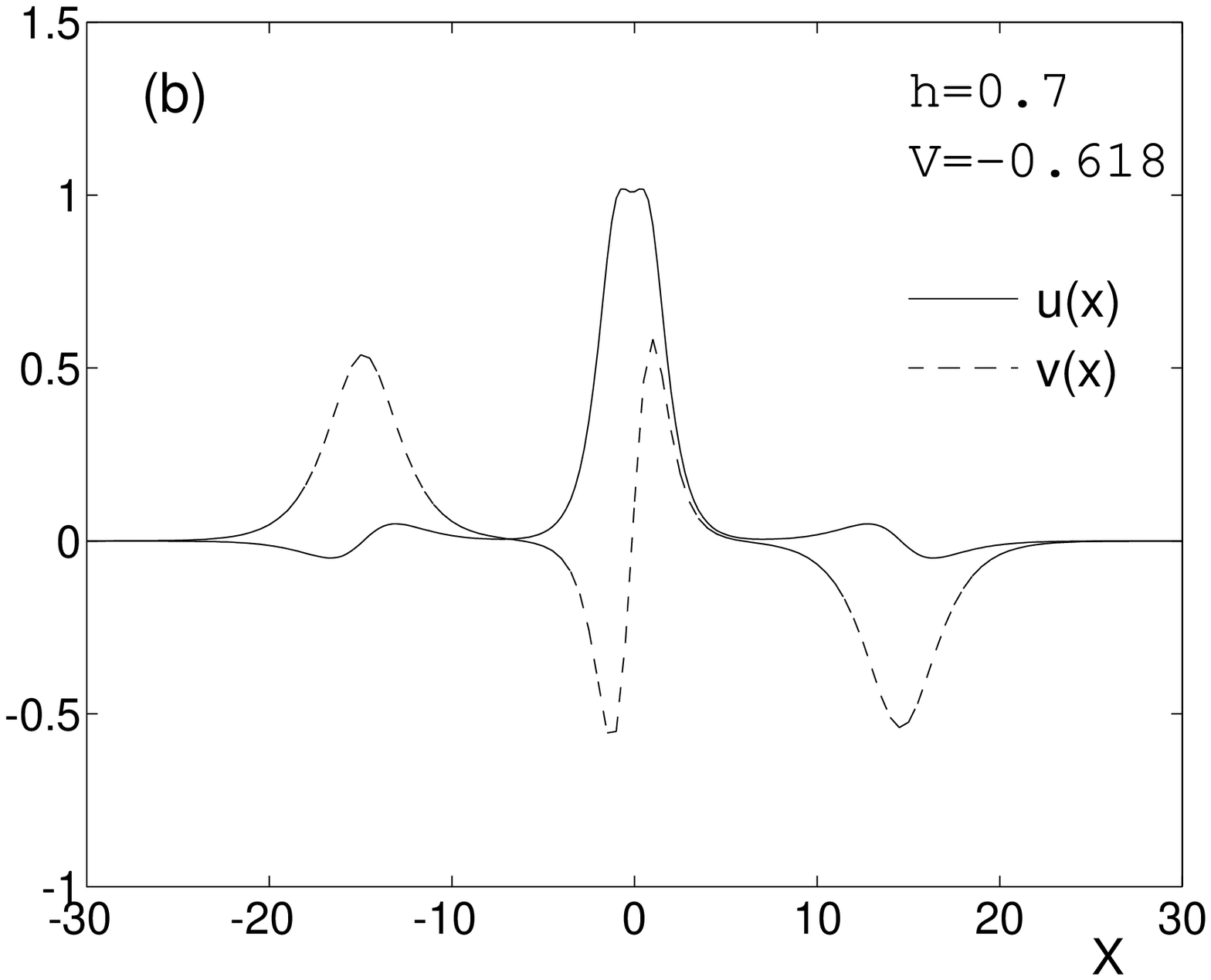,height=5cm,width=1.\linewidth}
\psfig{file=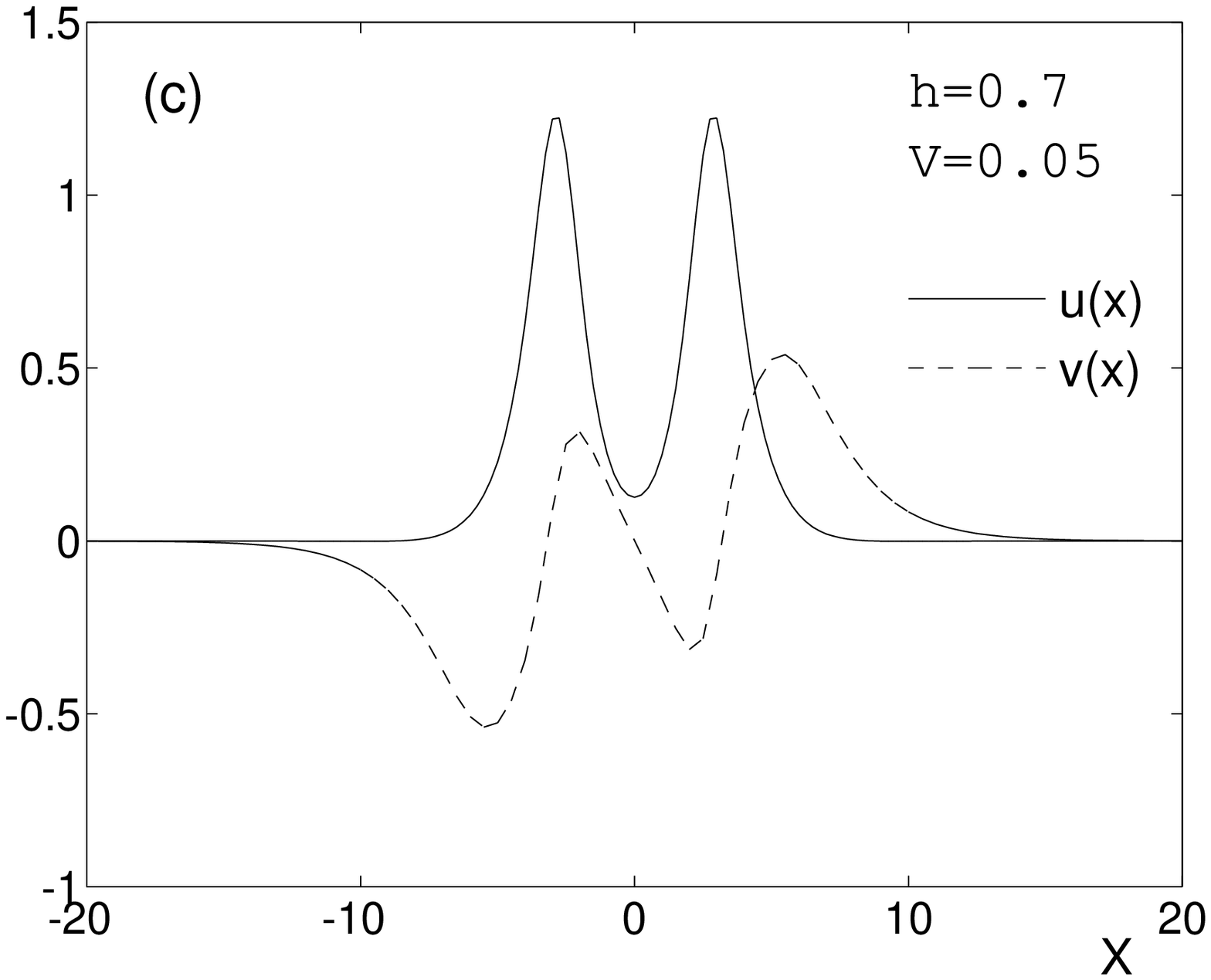,height=5cm,width=1.\linewidth}
\end{center}

\vspace{-12mm}
\label{twist_complexes}
\end{figure}
{\sf {\bf Fig.5}
(a): The $\psi_{(TT)}$ complex.
(b): The $\psi_{(-T-)}$ solution.
(c): A complex of two twist solitons arising from the
continuation of the $\psi_{(++)}$ bound state.
(Note the difference from the other two-twist complex
shown in (a).)
Solid line: real part; dashed line:
imaginary part}
\vspace{5mm}

Another branch emanating from the origin in Fig.4 (a), is
a bound state of two solitons $\psi_+$.
This
solution was {\it not\/} obtained  by the continuation
from $V=0$ as Fig.4  may seem to be suggesting.
Instead, we fixed a nonzero $V$ and continued
in $h$  from the value $h=0.05$ where the complex
$\psi_{(++)}$ arises from the $V$-continuation of the twist
soliton (see section \ref{further_le_28}).
Omitting details
of this procedure, we start the  description of the resulting branch
 at some
point $(V,P)$ away from the origin.
 As we approach the origin from this point, the
separation between the solitons $\psi_+$ in the complex
$\psi_{(++)}$
 rapidly increases
so that the field values
between the two solitons become exponentially small.
For example, for $h=0.7$,
 the (numerically calculated)
separation at the point $V=0$ was
equal to $z \approx 21$. The value
of $|\psi|$ at the point
on the $x$-axis,
 equally distanced from the left and right soliton,
was of order $10^{-6}$. Consequently, the nonlinear term in the
equation (\ref{stationary_NLS}) becomes  negligible away from the
solitons' core and, in spite of an extremely small value of
the  residual that we
used in our
numerical algorithm ($10^{-10}$),  we were unable to distinguish
between a genuine bound state and a linear superposition of
two distant solitons. We {\it conjecture\/} that the complex
$\psi_{(++)}$ exists all way to $V=0$ but as $V \to 0$, the
intersoliton separation $z \to \infty$. Another indication to this
effect is that as $V \to 0$, the imaginary part of the solution tends to
zero, rapidly and uniformly. For example, the imaginary part
of the above-mentioned numerical solution with the
real part
between the solitons
$ \geq 10^{-6}$, was smaller than
$10^{-13}$ for all $x$. Since
the only pure real solution that exists for $V=0$ is the (single)
soliton $\psi_+$, the $V \to 0$ limit of the $\psi_{(++)}$ complex
should be an infinitely separated pair of the $\psi_+$'s.

If we, conversely, continue our solution
 away from the origin, the curve $P(V)$ turns left at some $V$
and the complex $\psi_{(++)}$ transforms into what can be interpreted as
a bound state of two twists (denoted $\psi_{(TT)}$ in Fig.4.)
This solution is depicted in Fig.5(c). As $V \to 0$, the momentum
of this bound state tends to zero (Fig.4 (a)).
 Unfortunately, we were only able to obtain
this solution away from some small neighbourhood of $V=0$.
(For $h=0.7$, the smallest value of the velocity
for which we were still able to find the solution in question, was
$V=0.000283$.)   Whether this branch can be continued to $V=0$,
remains an open question.

\subsection{Other branches; $h <0.28$}
\label{further_le_28}

As we have mentioned, for $h <0.28$ the branch
$\psi_+$ extends all way to $V=c$ where
it merges with the zero solution.
No other solutions can be obtained from the  $\psi_+$ soliton.
However, in this case
we can obtain new
branches by continuing
the (quiescent) twist soliton, Eq.(\ref{complex})
with $z=\zeta$.

It is convenient to start our description with the motionless twist
solution with the {\it negative\/} momentum.
 As we move in the direction of
 positive $V$, the twist  gradually
transforms into a bound state of two $\psi_+$-solitons
(see Fig.6 (a)).
At some $V=V_{max}$ the branch turns back, shortly after which,
at the point $V=V_c$, the
momentum reaches its maximum and starts decreasing. Adjacent to
the turning point is a small range of velocities $V_c \le V \le V_{max}$
where we have two
{\it stable\/} solutions corresponding to each $V$.
If we continue the branch with {\it positive\/}-momentum twist solution,
also in the direction of positive $V$, the solution gradually
transforms into a complex of two $\psi_-$ solitons. The momentum
reaches its maximum, starts decreasing,
 then the branch turns back in $V$
and we find ourselves approaching the origin on the $(V,P)$-plane
(Fig.6 (a)). As we move towards the origin along the $\psi_{(++)}$
or along the $\psi_{(--)}$ branch, the separation between
 two solitons constituing the corresponding complex grows
while the imaginary part of the solution tends to zero.
Similarly to what we had for larger $h$
(section \ref{further_ge_28}), we conjecture that  the
separation becomes infinite at $V=0$ in both cases.

\begin{figure}
\begin{center}
\psfig{file=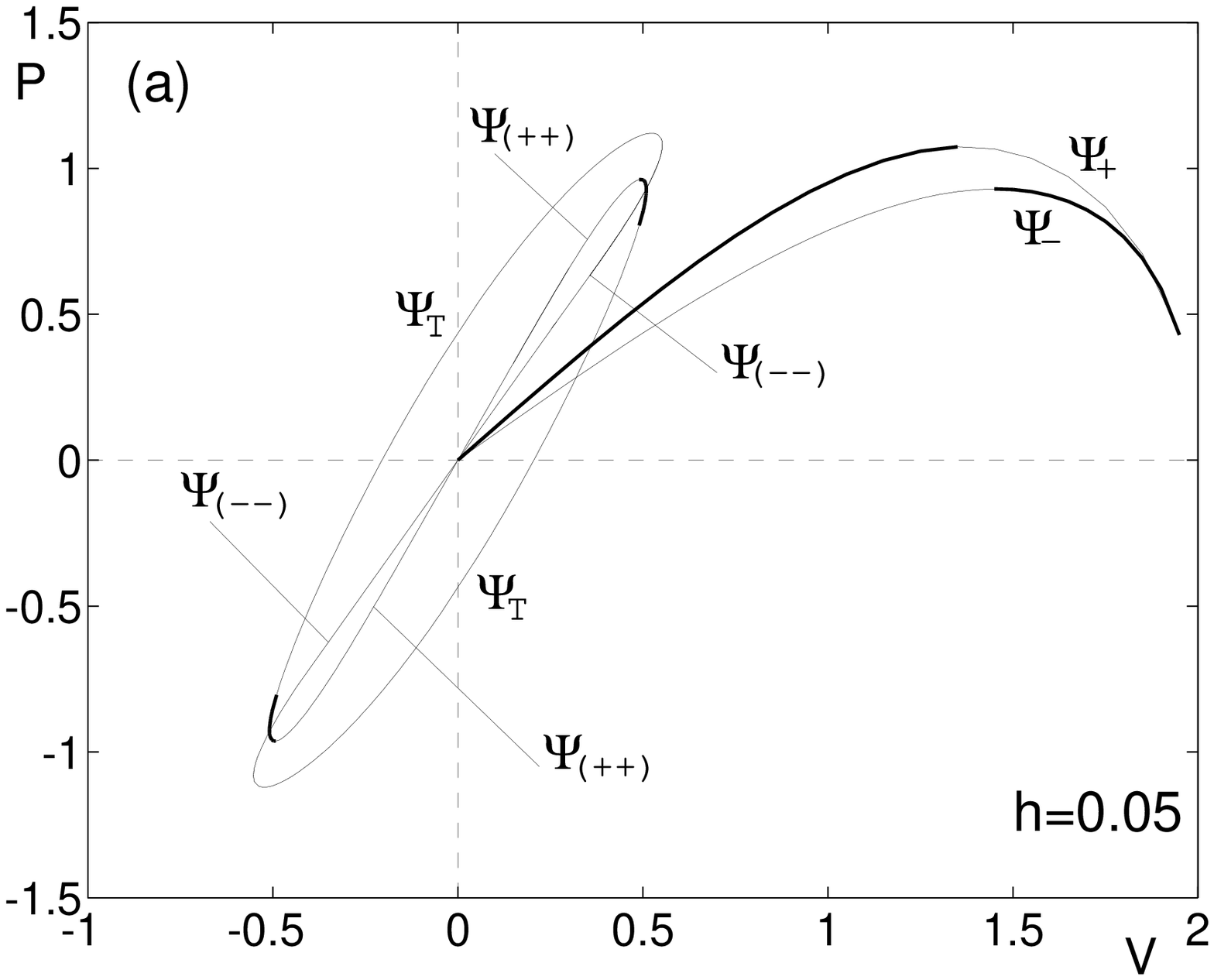,height=5cm,width=1.\linewidth}
\psfig{file=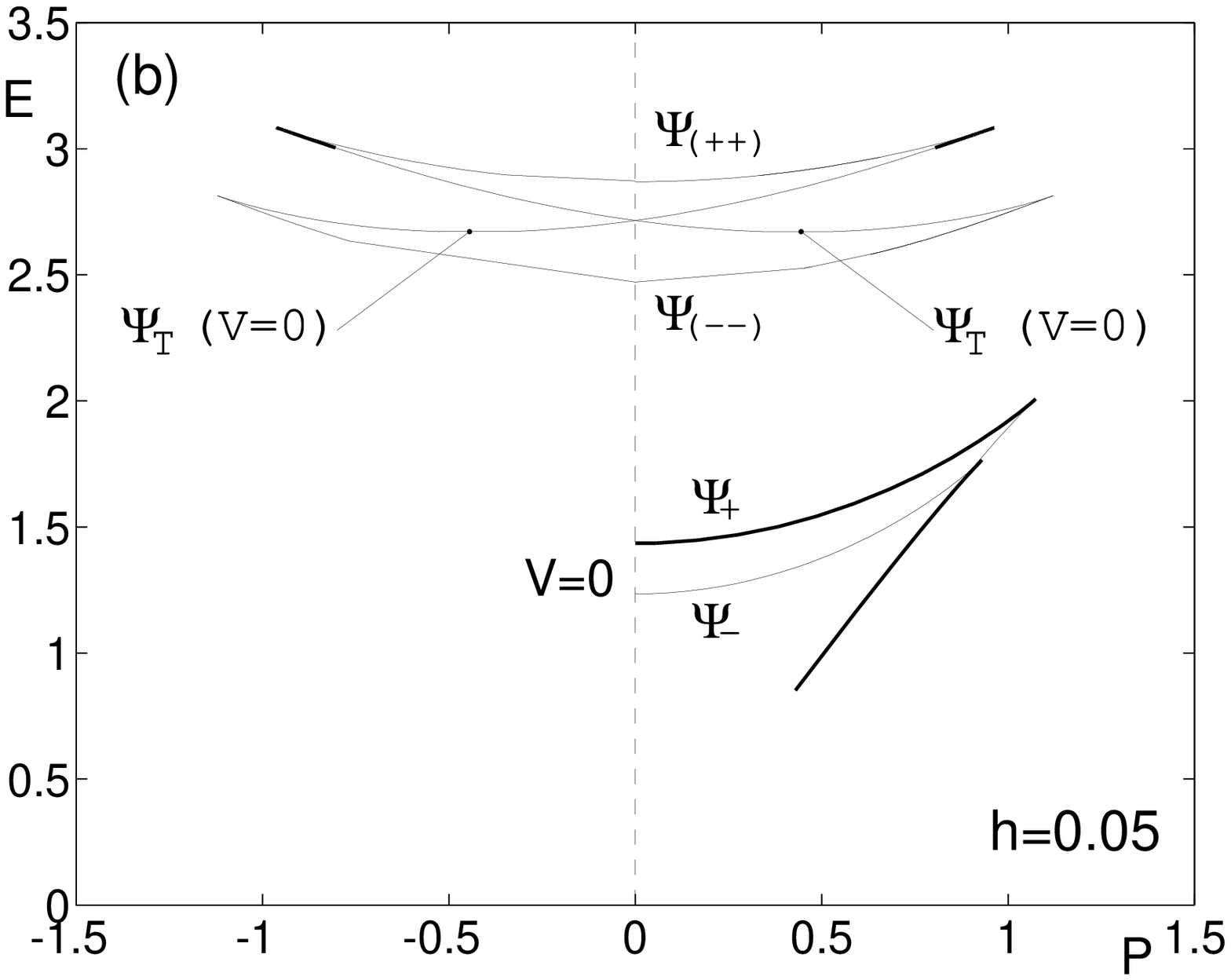,height=5cm,width=1.\linewidth}
\end{center}

\vspace{-12mm}
\label{bif_diagram}
\end{figure}
{\sf {\bf Fig.6}
(a): The full bifurcation diagram for $h=0.05$.
Thick respectively thin lines depict stable
respectively unstable branches. Note a region of stability
of the complex $\psi_{(++)}$.
Additional branches can be generated by
employing the reflection
symmetry $V \to -V$, $P \to -P$.
(b): The corresponding $E(P)$ diagram.
}
\vspace{5mm}

Fig.6 (b) shows the corresponding $E(P)$ dependence.
As in Fig.4 (b), the cusps mark  points of the zero crossing by
the stability eigenvalues. Note that as in the case of large $h$
(Fig.4 (b)),
there are  solitons with the same value of the momentum but different
energies. Similarly to Fig.4 (b), the stable branch of $\psi_-$ has the
lowest energy;  bound states on the stable $\psi_T \to \psi_{(++)}$
branch also have  lower
energies than their counterparts with the same $P$ and smaller $|V|$.
However, in the case of the $\psi_+$ solitons we have an interesting
reverse of fortunes: out of the two branches with the same $P$,
the stable branch is the one with {\it higher\/} energy!

\section{Nonlinear Stage of Instability}
\label{simulations}

 In this section we present results of our numerical simulations
of  the full time-dependent
nonlinear Schr\"odinger equation (\ref{NLS})
(with $\gamma=0$.)
The objective was to study the nonlinear stage of the development
of
instabilities reported in
the previous section and to identify the attractors emerging as
$t \to \infty$.
We utilised a split-step pseudospectral method,
with $2^{11}=2048$
modes on the intervals $-40 \leq X \leq 40$ and $-80 \leq X
\leq 80$,
and with
 $2^{12}=4096$ modes on the interval
$(-60,60)$.
The method imposes periodic boundary conditions
$\psi(L/2,t)=\psi(-L/2,t)$,
$\psi_X(L/2,t)=\psi_X(-L/2,t)$.

We have simulated  the evolution of moving solitons unstable
 against an oscillatory  mode and those with a positive,
nonoscillatory,  eigenvalue in their linearized spectrum.
One of our conclusions here is that
both types of instability give rise to the same asymptotic
attractors. (This is in agreement with earlier simulations
of motionless solitons \cite{Alex}.)

\subsection{The decaying breather}
Depending on the value of the
driving strength, the initial conditions and
the choice of   the parameters
of the numerical  scheme, we observed one of the two scenarios.
In the first scenario
the soliton transforms into a bell-shaped structure,
with a small amplitude and large spatial width,
oscillating approximately as $\psi \sim e^{i \omega t}$, with
{\it negative\/} $\omega$.
This localised solution was previously encountered in
numerical simulations of Ref.\cite{Alex}
where it was termed {\it breather\/}.
The amplitude  of the breather  slowly decays with time
and the width slowly grows.

We have detected this scenario for  the driving
strength $h=0.1$, with the initial condition in the form of the
$\psi_+$ soliton travelling  with the velocity
 $V=0.05$  and with $V=0.8$.
 (For both values of the velocity
the $\psi_+$ soliton is unstable against an oscillatory mode.)
Unlike earlier simulations  \cite{Alex} which started
with the  initial condition in the form of a   quiescent
unstable soliton  and gave
rise to a quiescent breather, the breather emerging from a
travelling soliton has a nonzero speed.

One may naturally wonder whether the speed of
the breather will decay to zero or approach a  nonzero constant value
as $t$ increases.
Our simulations seem to support the latter hypothesis.
In one run,
the speed of the breather evolving out of the
 soliton travelling with the initial velocity of $V=0.05$, was seen
to slowly grow and
 gradually approach the constant value of $0.1$.
This simulation was repeated, with the
same parameters of the numerical scheme
and an initial condition  which was only different from the
previous one due to interpolation errors of order $10^{-6}$.
In this run the breather was first seen to slow down, stop
but then start moving in the opposite direction with the
velocity close to -0.2; see Fig.7 (a).
(This remarkable sensitivity to the initial data deserves a
separate comment; see below.)
The velocity of the breather
evolving out of the $V=0.8$ soliton, was tending to
approximately $2.1$. However, for large
$t$ the unambigous interpretation
of the numerical data is hindered by the growth of the
amplitude of the radiation background. The radiation
waves emitted by the oscillating breather
re-enter the interval via the periodic boundary
conditions and  at a certain stage their amplitudes
become comparable with the amplitude of the breather.
Consequently, the constant-velocity motion of the
breather may have been induced by the interaction with
the backround radiations.

\begin{figure}
\begin{center}
\psfig{file=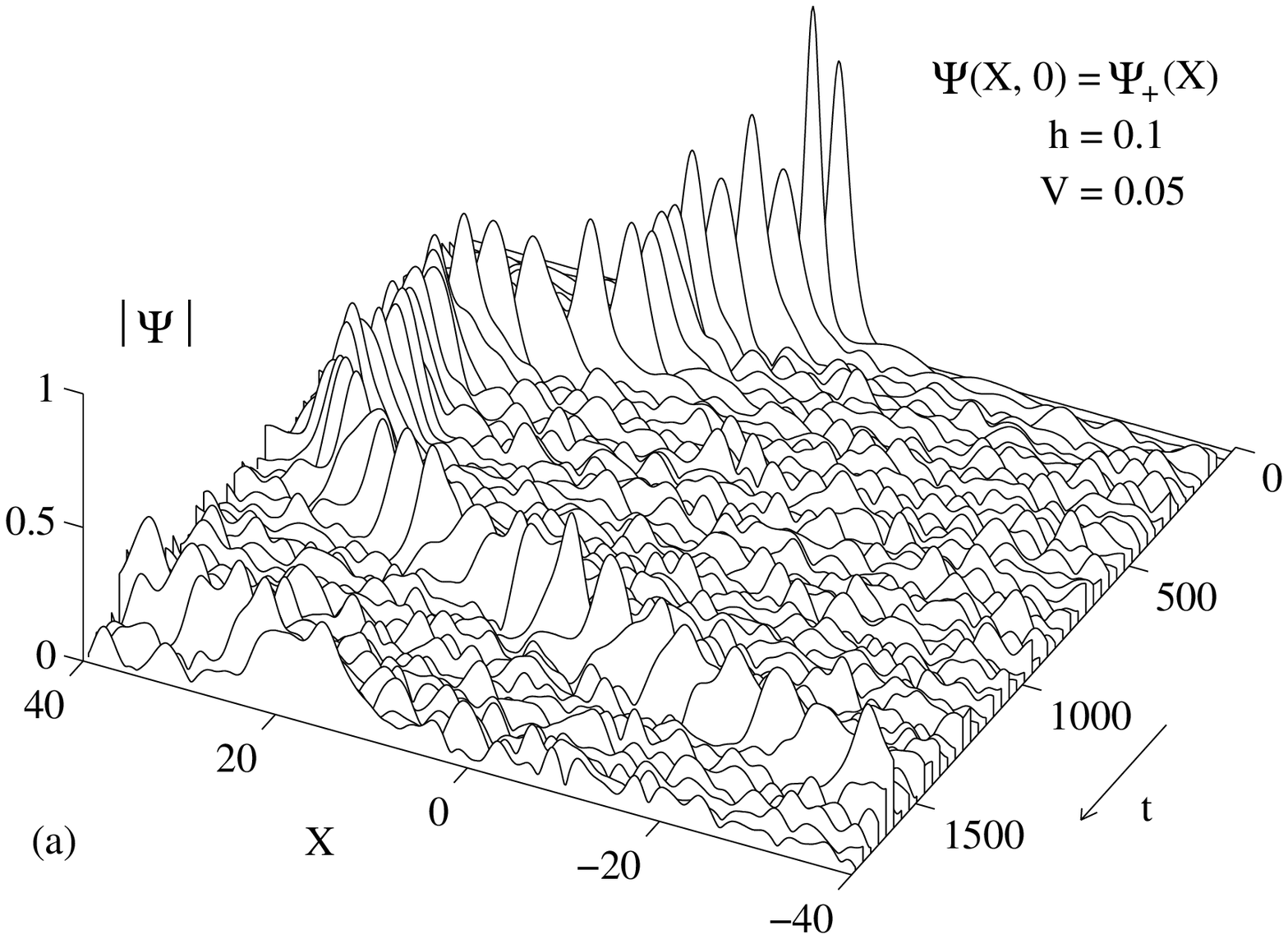,height=7cm,width=1.\linewidth}
\psfig{file=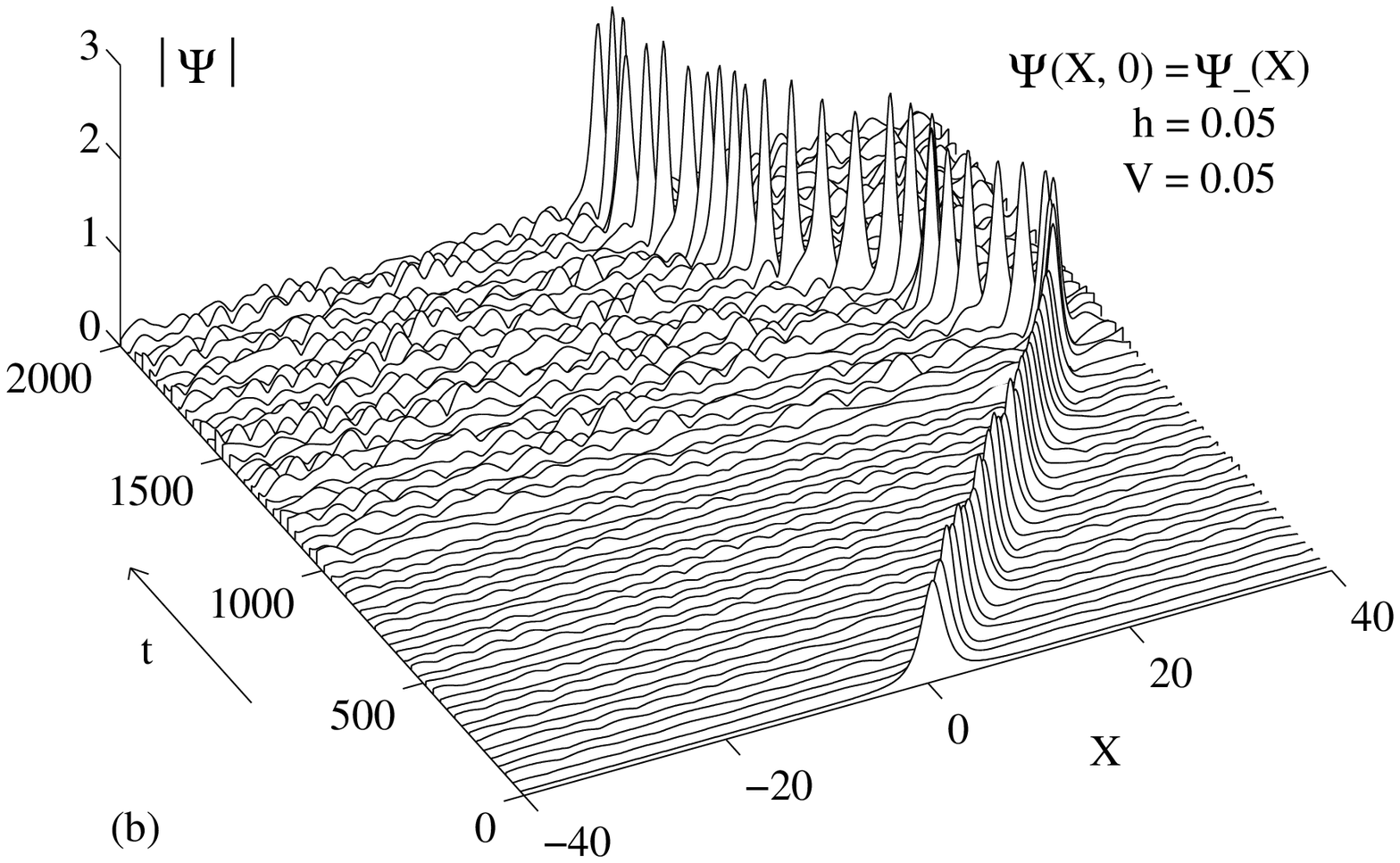,height=7cm,width=1.\linewidth}
\end{center}

\vspace{-10mm}
\label{3D}
\end{figure}
{\sf {\bf Fig.7}  The two types
of asymptotic attractors resulting from the
decay of the unstable steadily travelling solitons:
(a) the decaying and (b) the growing breather.
(a) corresponds to $h=0.1$ and the initial condition in the form of the
$\psi_+$ soliton with $V=0.05$. In (b),  $h=0.05$ and the
initial condition was chosen as the $\psi_-$ soliton with $V=0.05$.
In both plots the emerging breather changes, spontaneously,
its direction of motion. (Note that this happens
{\it not\/} as a result of the reflection from
 the boundary, as the periodic boundary conditions
are imposed.)
}

\vspace{4mm}

\subsection{The growing breather}
The decaying breather was detected in simulations  on
 the interval $(-40,40)$  with $N=2^{11}$ modes.
Changing the numerical parameters produced an entirely
different scenario, however, --- for the {\it same\/}
   value of the control parameter in Eq.(\ref{NLS})
($h=0.1$), and for the {\it same\/}
initial conditions ($V=0.05$ and $V=0.8$).

Namely, we increased the number of the Fourier modes to
$N=2^{12}$  and the length of the interval first to
$L=120$ and then to $160$.
 As in the case of $L=80$ and $N=2^{11}$, in
simulations with the new values of $N$ and $L$
 the unstable travelling soliton $\psi_+$  was seen to transform into
a bell-shaped  structure,
oscillating  roughly as
$\psi \sim e^{i \omega t}$.
 However,  this time the
emerging breather has a {\it positive \/}
frequency $\omega$; its amplitude is large and
continues to slowly grow, while the width is  narrow
and keeps on decreasing (Fig.7 (b)).

This attractor was also observed previously in \cite{Alex}.
It was found there that the decaying and growing breather
coexist.
Whether the evolution of the same unstable soliton
settles to one or the other asymptotic attractor, was found to depend
on the choice of the phase
  of a small perturbation
applied to  the  initial condition.
 In our present simulations,
 the  perturbation is introduced  simply by
 changing the parameters of the numerical scheme.

We  also examined initial conditions in the form of
{\it translationally\/} unstable solitons,
 including
the $\psi_+$-soliton with $V=1.4$ for the driving strength
$h=0.1$  and the
$\psi_-$-soliton with
initial velocities $V=0.05$
and $V=1.4$, for the driving strength $h=0.05$.
For each of the above three situations the
simulations were repeated with $2^{11}$ modes on the
interval $-40 \leq X \leq 40$, and with $2^{12}$ modes on the
intervals $(-60,60)$ and $(-80,80)$.
In all nine runs the unstable soliton was seen to evolve into the
growing breather.
(It is quite likely that some other choices of the numerical
parameters may give rise to the {\it decaying\/} breather instead.)

The velocity of the growing breather may vary during its
evolution. It can even wander erratically, changing the
direction of its motion several times, but
eventually, for $t \sim 10^4$ or even earlier, the  speed
of the breather locks on to some constant value.
Since the amplitudes of radiation waves are comparable
with the amplitude of the breather at that stage,
this effect can be induced by the breather-radiation interactions.

\section{Concluding remarks and open problems}
\label{conclusions}

The main result of  this paper is the demonstration of the
existence of wide classes of travelling soliton solutions
of the (undamped) parametrically driven nonlinear Schr\"odinger
equation. We  established the necessary conditions under
which motionless solitons can be continued to nonzero velocities,
and, in cases where these conditions were met, were indeed
able to carry out the numerical continuation.
The stability of all resulting branches of solutions was
 examined;
  oscillatory and translational instabilities
identified, and
 the single-soliton stability chart compiled on the $(h,V)$-plane.
The onset values  of the translational instabilities,
obtained numerically, were
shown to verify the relation $\partial P/\partial V=0$
predicted by our theoretical analysis.
As opposed to the case of  the soliton $\psi_-$, which undergoes
similar transformations for any $h$,
the result of the continuation of the
$\psi_+$ has turned out to be sensitive to the value
of the driving strength.
We have identified
two different transformation scenarios,
one occurring for small  and the other one for larger $h$,
 and
described an interesting cross-over from one to the other.

In our analysis we paid a special attention to the trajectories
of linearised eigenvalues on the complex plane.
Apart from the information on the stability of different branches
of solutions,  the behaviour of the
eigenvalues can give insight into the {\it supercritical\/} dynamics
of solitons, i.e. dynamics beyond the instability threshold \cite{Alex}.
The motion of the eigenvalues in the undamped case
that we are currently concerned with, allows one to even predict
the asymptotic attractors arising when there is a small but nonzero
damping in the system \cite{Alex}. Relegating the corresponding
 bifurcation
analysis
to  future publications, here
we have restricted ourselves to a series of
  numerical simulations of the
time-dependent NLS equation (\ref{NLS}) (with $\gamma=0$.)

It is worth listing, separately, {\it stable\/} solutions obtained in
this study.
First, the quiescent soliton $\psi_-$, which
is always
unstable with respect to a nonoscillatory
mode, stabilizes when travels faster than a certain
critical velocity. The stability boundary satisfies
$\partial P/\partial V=0$.
Second, the quiscent soliton  $\psi_+$ is stable for
$h<0.064$ and loses its stability to an oscillating
soliton for $h>0.064$. For nonzero $V$ the stability region
is shown in Fig.3. The lower boundary of this
 region corresponds to the onset of the oscillatory instability
while along the upper boundary the soliton becomes unstable
with respect to a nonoscillatory mode. The corresponding
critical velocity satisfies
$\partial P/\partial V=0$.
Third, the bound state $\psi_{(++)}$ also displays a
region of stability for small $h$ --- see Fig.6.

We conclude this section by pointing out to
several open questions.
In the first place, it would be interesting to
continue the $\psi_{(-T-)}$-branch in Fig.4. Will this multisoliton  complex
keep on attaching more solitons on its flanks?
Another ``open-ended" branch  ($\psi_{(TT)}$)
approaches the origin vertically down in Fig.4 (a);
it would be interesting to continue this branch as well.
 The striking difference between
the bifurcation diagrams for $h>0.28$ (Fig.4) and
$h<0.28$ (Fig.6) is also
worthy of a deeper analysis. Is the bifurcation diagram in
Fig.6 complete, or there are other multisoliton branches similar to those
arising for large $h$ (Fig.4)?
The next open question concerns the decaying and growing breather
solutions arising as a result of the growth of the instability of
steadily travelling solitons. If the radiations were prevented from
re-entering the interval of simulation, would the velocity
of the breather decay to zero or approach some nonzero value determined
by the initial velocity of the unstable soliton?
Finally, a
challenging problem is to find out whether travelling solitons
can exist in the presence of damping.
The effect of dissipation
will be, of course, to attenuate the soliton.
It is not obvious whether
 the spatially uniform
parametric pumping
is capable of compensating this type of losses and
sustaining the soliton's steady motion.
It is fitting to note here that the parametrically driven,
damped nonlinear
Schr\"odinger equation has wide classes of stationary solitons
\cite{more_solutions}, some of which may be continuable to nonzero $V$.

\acknowledgements
We thank Dmitry Pelinovsky for drawing our attention to
Ref.\cite{TS} and solution (\ref{complex}) presented therein.
Useful conversations
with Yuri Gaididei   are also  gratefully acknowledged.
Special thanks go to Nora Alexeeva
for providing us with a
code for the time-dependent NLS equation and
for her help with
numerics.
Finally, we are grateful to Professor Igor Puzynin
for his continual encouragement and strong administrative
support of this project.
 I.B. was  supported by the
NRF of South Africa and URC of the University of Cape Town.
E.Z. was  supported by an RFBR grant No.0001-00617.


\end{multicols}

\begin{thebibliography}{99}
\bibitem[\dag]{add:igor}
 On leave  from
Department of Mathematics,
University of Cape Town, Private Bag Rondebosch 7701, South Africa.
 Email:
 igor@maths.uct.ac.za
\bibitem[\S]{add:elena}
 On leave  from
the Laboratory for Computing Techniques,
Joint Institute for Nuclear Research, Dubna, 141980, Russia.
 Email: elena@ultra.jinr.ru
\bibitem[\ddag]{email:markus} Email: baer@mpipks-dresden.mpg.de

\bibitem{water}
J.W. Miles, J. Fluid Mech. {\bf 148}, 451 (1984);
M.  Umeki, J.  Phys.  Soc.  Jpn.  {\bf 60}, 146 (1991); J.
Fluid Mech.  {\bf 227}, 161 (1991)


\bibitem{water_exp}
X.N. Chen and R.J. Wei, J. Fluid Mech. {\bf 259}, 291 (1994);
X. Wang, R. Wei, Phys. Lett. A {\bf 192}, 1 (1994);
W. Wang, X. Wang, J. Wang, R. Wei, Phys. Lett. A {\bf 219}, 74 (1996);
X. Wang, R. Wei, Phys. Lett. A {\bf 227}, 55 (1997);
X. Wang and R. Wei,
Phys. Rev. Lett. {\bf 78}, 2744 (1997); Phys. Rev. E {\bf 57}, 2405 (1998);
X. Wang, J. Acoust. Soc. Am. {\bf 104}, 715 (1998)

\bibitem{optics} I.H.  Deutsch and I.  Abram, J.  Opt.  Soc.  Am.  {\bf B11},
2303 (1994);
  A.  Mecozzi, L. Kath,
P.  Kumar, and C.G.  Goedde, Opt.Lett.  {\bf 19}, 2050 (1994);
 S.  Longhi, Opt.Lett.  20, 695 (1995);
 Phys. Rev. {\bf E 55}, 1060 (1997)


\bibitem{magnetism} V.E.  Zakharov, V.S.  L'vov and S.S.  Starobinets, Sov. Phys.
Uspekhi {\bf 17}, 896 (1975);
M.M.  Bogdan, A.M.  Kosevich and I.V.  Manzhos, Sov.  J.  Low
Temp.  Phys.  {\bf 11}, 547 (1985);
  H.  Yamazaki and M.  Mino, Prog.  Theor.  Phys.  Suppl. {\bf
98}, 400 (1989)

\bibitem{BBK} I.V.  Barashenkov, M.M.  Bogdan, and V.I.  Korobov, Europhys.
Lett.  {\bf 15}, 113 (1991)

\bibitem{XY} L.N. Bulaevskii and V.L. Ginzburg, Sov. Phys. JETP
{\bf 18}, 530 (1964); J. Lajzerowicz and J.J. Niez,
 Journ. de Physique, {\bf 40}, L165 (1979)


\bibitem{Scott}
 A.V. Ustinov, B.A. Malomed, S. Sakai,
Phys. Rev. {\bf B 57}, 11691 (1998);
H.S.J. van der Zant {\it et al\/}, Physica {\bf D 119}, 219 (1998);
O.M. Braun, Yu.S. Kivshar, Phys. Rep. {\bf 306}, 1 (1998);
A.C. Scott, Nonlinear Science.
Emergence and Dynamics of Coherent Structures.
Oxford Univ. Press (1999).

\bibitem{Ginsburg_Landau}
P. Coullet, J. Lega, B. Houchmanzadeh, and J. Lajzerowicz,
Phys. Rev. Lett. {\bf 63}, 1352 (1990);
C. Elphick, A. Hagberg, B.A. Malomed, and E. Meron;
Phys. Lett. {\bf A 230}, 33 (1997)



\bibitem{dark_solitons}
I.V. Barashenkov and V.G. Makhankov, Phys. Lett. A {\bf 128},
52 (1988);
I.V. Barashenkov, T.L. Boyadjiev, I.V. Puzynin, and T. Zhanlav,
Phys. Lett. {\bf A 135}, 125 (1989);
M.M. Bogdan, A.S. Kovalev, and A.M. Kosevich,
Sov. J. Low Temp. Phys., {\bf 15}, 288 (1989);
I.V. Barashenkov and E.Yu. Panova, Physica D {\bf 69}, 114 (1993);
Yu.S. Kivshar and X. Yang, Phys. Rev. E {\bf 49}, 1657 (1994)

\bibitem{Bar} I.V. Barashenkov,
Phys. Rev. Lett. {\bf 77}, 1193 (1996)

\bibitem{PAK}
D.E. Pelinovsky, Yu.S. Kivshar, and V.V. Afanasjev,
Phys. Rev. {\bf E 54}, 2015 (1996)

\bibitem{Alex} N.V. Alexeeva, I.V. Barashenkov and D.E. Pelinovsky,
Nonlinearity {\bf 12}, 103 (1999).




\bibitem{TS} M.V. Tratnik and J.E. Sipe,
Phys. Rev. {\bf A 38}, 2011 (1988)


\bibitem{more_solutions}
 I.V. Barashenkov and E.V. Zemlyanaya,
Phys. Rev. Lett. {\bf 83}, 2568 (1999);
A. Il'ichev, Physica D {\bf 119}, 327 (1998);
I. Bakholdin, A. Il'ichev, Eur. J. Mech. B/Fluids, {\bf 18},
93 (1999)




\end{thebibliography}
\end{document}